\documentclass[11pt,a4paper,onecolumn,reqno]{amsart}
\usepackage[a4paper, margin=2.3cm, bmargin=3cm]{geometry}

\usepackage{graphicx,stfloats} \usepackage{color}
\usepackage{amsmath}
\usepackage{amsaddr}
\usepackage{bm, soul}
\usepackage{mathtools}
\usepackage[colorinlistoftodos,prependcaption,textsize=tiny]{todonotes}
\usepackage{braket}
\usepackage{hyperref}
\usepackage[square,comma,sort&compress, numbers]{natbib}
\usepackage{enumerate}
\usepackage{enumitem}
\usepackage{floatrow}
\usepackage{tikz}
\usepackage{pgfplots}
\usepackage{csvsimple}
\usepackage{pict2e}
\usepackage{overpic}
\pgfplotsset{compat=1.9}
\usetikzlibrary{calc,arrows,arrows.meta,fadings,decorations.pathreplacing,decorations.markings,patterns,shapes.geometric}
\tikzset{>=stealth,inner sep=0pt, outer sep=2pt,}
\tikzset{vecteur/.style={->,thick,color=black,smooth}}

\usepackage{placeins}

\renewcommand{\st}[1]{}

\usepackage[version=3]{mhchem} \usepackage{siunitx}
\DeclareSIUnit\bar{bar}
 \usepackage{multirow}

\usepackage{etoolbox}
\patchcmd{\pprintMaketitle}
  {\hrule\vskip12pt}
 {\hrule\vskip12pt\ifvoid\extrainfobox\else\unvbox\extrainfobox\par\vskip12pt\fi}
 {}{}

\newsavebox\extrainfobox

\usepackage{caption}
\title{
    Flame-wall interaction of thermodiffusively unstable hydrogen/air flames -- Part II: Parametric variations of equivalence ratio, temperature, and pressure
}

\author[stfs]{Max Schneider$^{*}$, Hendrik Nicolai, Vinzenz Schuh, Matthias Steinhausen, Christian Hasse}
\email{schneider@stfs.tu-darmstadt.de} 
\address[]{Technical University of Darmstadt, Department of Mechanical Engineering, Simulation of reactive Thermo-Fluid Systems, Otto-Berndt-Stra{\ss}e 2, 64287 Darmstadt, Germany\\
}

\begin{document}
\pagestyle{plain}

\maketitle

\begin{abstract} 
Fuel-lean hydrogen combustion systems hold significant potential for low pollutant emissions, but are also susceptible to intrinsic combustion instabilities.
While most research on these instabilities has focused on flames without wall confinement, practical combustors are typically enclosed by walls that strongly influence the combustion dynamics.
In part I of this work \cite{PartI}, the flame-wall interaction of intrinsically unstable hydrogen/air flames has been studied for a single operating condition through detailed numerical simulations in a two-dimensional head-on quenching configuration.
This study extends the previous investigation to a wide range of gas turbine and engine-relevant operating conditions, including variations in equivalence ratio ($0.4$ - $1.0$), unburnt gas temperature ($\SI{298}{\kelvin}$ - $\SI{700}{\kelvin}$), and pressure ($\SI{1.01325}{\bar}$ - $\SI{20}{\bar}$).
These parametric variations allow for a detailed analysis and establish a baseline for modeling the effects of varying instability intensities on the quenching process, as the relative influence of thermodiffusive and hydrodynamic instabilities depends on the operating conditions.
While the quenching characteristics remain largely unaffected by hydrodynamic instabilities, the presence of thermodiffusive instabilities significantly increases the mean wall-heat flux and reduces the mean quenching distance.
Furthermore, the impact of thermodiffusive instabilities on the quenching process intensifies as their intensity increases, driven by an increase in pressures and a decrease in equivalence ratio and unburnt gas temperature.
The corresponding relative increase in wall heat flux, compared to a one-dimensional stable head-on quenching flame under identical operating conditions, strongly correlates with the enhanced local reactivity associated with thermodiffusive instabilities across all operating conditions.
Finally, a joint model fit is proposed to estimate the relative increase in wall heat flux, with local reactivity serving as an intermediate step, based on characteristic quantities of a corresponding one-dimensional stable freely-propagating flame.
\end{abstract}


\section*{Novelty and Significance Statement}
This work presents a novel parametric study of flame-wall interactions (head-on quenching) of intrinsically unstable hydrogen/air flames. 
It builds upon a previous investigation of an unstable head-on quenching hydrogen/air flame (part I) and extends it to a wide range of operation conditions, including variations in equivalence ratio, unburnt gas temperature, and pressure.
The study shows the influence of thermodiffusive and hydrodynamic instabilities on the quenching process by analyzing the local wall heat flux, as well as, global quenching characteristics.
Furthermore, the study reveals that the intensity of thermodiffusive instabilities correlates well with the increase in the peak wall heat flux relative to a one-dimensional simulation for all operating conditions investigated.
To enable the assessment of thermal loads in technical applications, a novel model fit is proposed that allows to estimate the peak wall heat flux during quenching based on characteristic quantities of one-dimensional flames.

\section*{CRediT authorship contribution statement}
\textbf{Max Schneider}: Conceptualization, Methodology, Investigation, Software, Formal analysis, Data Curation, Visualization, Writing - original draft.
\textbf{Hendrik Nicolai}: Conceptualization, Methodology, Supervision, Project administration, Writing - review \& editing.
\textbf{Vinzenz Schuh}: Software, Formal analysis, Writing - review \& editing.
\textbf{Matthias Steinhausen}: Conceptualization, Methodology, Supervision, Writing - review \& editing.
\textbf{Christian Hasse}: Resources, Supervision, Project administration, Funding acquisition, Writing - review \& editing.

\section{Introduction} \addvspace{10pt}
\label{sec:introduction}
\setlist{nolistsep}

Fuel-lean premixed hydrogen/air flames are preferred in technical applications due to their lower \ce{NOx} emissions compared to other combustion strategies. 
However, they are prone to intrinsic combustion instabilities, which can substantially alter flame dynamics, heat release rates, and also influence flame-wall interaction (FWI)~\cite{PartI}, all of which are critical for safety and thermal efficiency \cite{Verhelst2009}.
The intrinsic combustion instabilities are caused by the superposition of thermodiffusive and hydrodynamic instabilities.
Thermodiffusive instabilities are associated with hydrogen’s low Lewis number, a consequence of its high molecular diffusivity.
Hydrodynamic instabilities arise due to the density jump across the flame front, a characteristic feature of all premixed flames.

The aforementioned FWI is a consequence of the confinement of a flame by solid walls in a technical combustion chamber, such as in gas turbines and internal combustion (IC) engines, and therefore an important research topic in technical combustion applications.
As the flame approaches the wall, it induces a significant wall heat flux, which weakens the flame.
This interaction leads to material aging due to the significant heat loads on the combustor walls, decreases combustion efficiency, affects pollutant formation \cite{Dreizler2015, Lai2018, Steinhausen2023}, and may result in undesirable flame behavior, such as flashback \cite{Fritz2004}.
Furthermore, current trends in engine design involve downsizing to increase the power density and enhance fuel efficiency \cite{Leach2020}.
However, these downsized combustors exhibit an increased surface-area-to-volume ratio, resulting in greater exposure of the flame to the walls, while higher pressures lead to intensified chemical reactions near surfaces \cite{Johe2022}.
FWI is particularly significant for hydrogen/air flames, as the high reactivity of \ce{H2} causes the flame to approach closer to the walls, thereby increasing wall heat fluxes and thermal loads compared to hydrocarbons \cite{Dabireau2003, Gruber2010, DeNardi2024}.
Although FWI, particularly in hydrogen/air flames, has been investigated in numerous works, as discussed in part I \cite{PartI} of this work, the additional impact of thermodiffusive instabilities on the quenching process remains unexplored.
Moreover, most studies on thermodiffusive instabilities are focused on non-enclosed domains, and consequently do not address FWI.

Hence, this numerical study aims to address the gap in research concerning FWI of thermodiffusively unstable flames.
In part I of this numerical study, the effects of thermodiffusive instabilities on FWI are analyzed and characterized for a single operating condition ($\varphi = 0.4$, $T_{\mathrm{u}} = \SI{298}{\kelvin}$, $p = \SI{1.01325}{\bar}$) in a two-dimensional head-on quenching (HOQ\footnote{HOQ is a transient process characterized by a flame front approaching a wall, commonly employed to study FWI~\cite{Dreizler2015}.}) setup.
A large numerical domain is used to avoid suppressing instability length scales, and additional insights are gained by employing setups with reduced lateral domain size to limit the maximum cell size of the instabilities.

Understanding the effects of intrinsic instabilities across a broad parametric space is essential.
Accordingly, this part employs parametric variations of the equivalence ratio $\varphi$, the unburnt temperature $T_{\mathrm{u}}$, and the pressure $p$ in a similar setup.
The operating conditions are adapted from a study by Berger et al.~\cite{Berger2022, Berger2022a}, who examined the impact of intrinsic combustion instabilities on lean hydrogen/air premixed freely-propagating flames across a range of equivalence ratios ($\varphi = 0.4 - 1$), unburnt temperatures ($T_{\mathrm{unburnt}} = \SI{298}{\kelvin} - \SI{700}{\kelvin}$), and pressures ($p = \SI{1}{bar} - \SI{20}{bar}$).
It is demonstrated that the impact of instabilities increases with a decreasing equivalence ratio, a decreasing unburnt gas temperature and an increasing pressure.
The increasing impact of the instabilities on the flame dynamics is illustrated by an increasing global consumption speed relative to the laminar burning velocity $s_{\mathrm{c}} / s_{\mathrm{l}}$, which results from both an increasing surface area $A / L_{y}$ and reactivity factor $I_0$, as also demonstrated in \cite{Altantzis2015, Frouzakis2015, Berger2019, Creta2020, Attili2021, Howarth2022}.
Furthermore, they propose joint scaling laws for the flame speed enhancement, expressed as a function of the equivalence ratio $\varphi$, the unburnt temperature $T_{\mathrm{u}}$, and the pressure $p$, as well as in terms of global flame parameters such as the expansion ratio $\sigma$, the Zeldovich number $Ze$, and the effective Lewis number ${Le}_{\mathrm{eff}}$.
Similar parametric variations (equivalence ratio $\varphi$, unburnt temperature $T_{\mathrm{u}}$ and pressure $p$) in smaller domains have been performed by Howarth et al.~\cite{Howarth2022, Howarth2023} by means of two-dimensional and three-dimensional direct numerical simulations of lean hydrogen/air flames.
They also propose empirical models for predicting both local and characteristic values of flame speeds and thicknesses for the two-dimensional and three-dimensional unstable flames, which can be assessed from one-dimensional simulations.
Rieth et al.~\cite{Rieth2023} performed parametric pressure variations in a canonical turbulent flame configuration within homogeneous isotropic turbulence.
They found that the ratio of turbulent to laminar burning velocity increases with pressure; however, this effect is not driven by pressure itself but rather by flame properties that scale with pressure, such as the Zeldovich number $Ze$ and a convection-diffusion Péclet number $Pe_{\mathrm{cd}}$, which describes the balance between convective and diffusive transport.
The authors also propose a power law to scale the reactivity factor $I_0$ based on the ratio of the Zeldovich and Péclet number $Ze/Pe_{\mathrm{cd}}$.

Building on these results, this study examines the effects of intrinsic combustion instabilities within a FWI configuration, rather than in a freely-propagating flame setup as in the works outlined above.
It aims to provide a comprehensive picture of the effects of intrinsic combustion instabilities on FWI for different conditions of practical interest, which are associated with different intensities of thermodiffusive instabilities.
Thus, direct numerical simulations (DNS) of a HOQ setup are conducted over a wide range of operating conditions, incorporating detailed transport and chemical kinetics.

The main objectives are to:
\begin{itemize}[noitemsep]
    \item examine the effects of varying operating conditions on the FWI in a one-dimensional head-on quenching configuration for stable flames to establish a baseline for the subsequent analysis,
    \item analyze the impact of intrinsic combustion instabilities on the FWI across the varying operating conditions,  
    \item investigate the contributions of thermodiffusive and hydrodynamic instabilities on the quenching process, and,
    \item derive an approximation of the increased heat load on the combustor wall resulting from instabilities as a function of characteristic quantities of a one-dimensional freely-propagating flame.  
\end{itemize}

The outline of the paper is as follows:
Section \ref{sec:NumericalSetup} provides a detailed description of the numerical setup for the simulations of one-dimensional HOQ of stable flames, that serve as reference, and the two-dimensional HOQ of the unstable flames.
The subsequent Sec. \ref{sec:1D} highlights the trends in global flame and quenching parameters in one-dimensional freely-propagating flames and one-dimensional HOQs, such as flame power and peak wall heat flux, that result from variations in the equivalence ratio, unburnt temperature, and pressure.
These trends are essential for analyzing and characterizing the trends for similar variations in the two-dimensional HOQ configuration in Sec. \ref{sec:2DHOQ}, where a model is also proposed to approximate wall heat flux enhancement due to intrinsic combustion instabilities.
Finally, the main conclusions are presented in Sec. \ref{sec:Conclusions}.

\section{Numerical setup} \addvspace{10pt}
\label{sec:NumericalSetup}
The numerical configuration for the one-dimensional HOQ of stable flames and the two-dimensional HOQ of thermodiffusively unstable flames is detailed below.
Following this, the operating conditions for the parametric variations are described.
Finally, the numerical methods employed in this study are briefly outlined.

\subsection{Configuration}

\subsubsection{One-dimensional head-on quenching}
\begin{figure}[h!]
    \centering
    \begin{tikzpicture}[scale=1.4]
        \draw[smooth, samples=800, domain=0:0.5, color=orange] plot({-0.55}, \x);
        \draw [orange, ->](-0.55, 0.25) -- (-0.85,0.25);
        
        \draw[black, thick] (-2,0) rectangle (2,0.5);
        \node [] at (-1.4, 0.25) { $Y_{\mathrm{u}}$, $T_{\mathrm{u}}$};
        \node [] at (0.5, 0.25) { $Y_{\mathrm{b}}$, $T_{\mathrm{b}}$};
        \draw[thick](-2.2,0)--++(0.3,0);
        \draw[thick](-2.2,0.5)--++(0.3,0);
        \fill[pattern=north west lines] (-2.2,0) rectangle (-2, 0.5);
        \draw[thick](-2,0)--++(0,-0.3);
        \draw [-{Stealth}](-2, -0.2) -- (-1.75,-0.2);
        \node [] at (-1.65, -0.28) { $x$};

        \node[color=gray] at (-1.9, 0.75) { wall};
        \node[color=gray] at (1.55, 0.75) { outlet};
    \end{tikzpicture}
    \caption{\footnotesize
        Schematic of the one-dimensional HOQ configuration.
        An exemplary flame front is highlighted by the orange line.
    }
    \label{fig:configuration_1D_HOQ}
\end{figure}
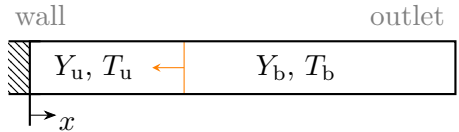

Figure \ref{fig:configuration_1D_HOQ} shows the configuration of the one-dimensional HOQ.
In streamwise direction, a wall is on the left side and an outlet is on the right side.
At the wall, a zero-flux boundary condition is imposed for the species, ensuring zero mass flux for all species across the wall (for details, see Part I \cite{PartI}), and a no-slip condition is enforced for the velocity.
The wall temperature $T_{\mathrm{wall}}$ is set equal to the respective unburnt gas temperature $T_{\mathrm{u}}$.
At the outlet, a zero-gradient boundary condition is specified for the reactive scalars and the velocity.
Initially, the fresh gases are at rest and the flame moves towards the wall at laminar flame speed $s_{\mathrm{l}}$, with the initial flame profile computed with Cantera \cite{cantera}. The computations are conducted using the well-established mechanism by Li et al.~\cite{Li2004}.

\subsubsection{Two-dimensional head-on quenching}

The setup for the two-dimensional HOQ of the thermodiffusively unstable flame is similar to part I of this work \cite{PartI}.
The computational domain is shown in Fig. \ref{fig:configuration_2D_HOQ}.
The boundary conditions are identical to those of the one-dimensional HOQ configuration, with periodic boundary conditions applied in the $y$-direction at the top and bottom of the domain.
The simulations are initialized with temperature, velocity, and species profiles from initial simulations of unstable hydrogen/air flames in the nonlinear regime, shown in Fig. \ref{fig:configuration_initial}.
For the initial simulations the boundary conditions are similar to those of the two-dimensional HOQ, except for the boundary conditions at the inlet on the left side of the domain.
At the inlet, a hydrogen/air mixture enters the domain.
For the velocity, a boundary condition is employed that stabilizes the flame in the center of the domain (for further details, see part I \cite{PartI}).
The initial freely-propagating flame simulations are initialized using one-dimensional freely-propagating unstretched flames from Cantera \cite{cantera}, aligned in $x$-direction and perturbed with a weak initial sinusoidal perturbation, as illustrated in Fig. \ref{fig:configuration_initial}.
Simulations were run for over $100$ flame times $\tau_T^0$ to ensure that the flame propagates within the non-linear regime, the flame dynamics are statistically stationary, and the time-averaged consumption speed remains constant.
It is important to note that the amplitude and wavelength of the perturbation do not affect the non-linear regime \cite{Berger2019}.

The equivalence ratio $\varphi$ and the unburnt temperature of the mixture $T_{\mathrm{u}}$, as well as the global pressure $p$ within the domain are varied in this study, which is discussed in detail below.

The dimensions $L_y$ and $L_x$ are expressed as multiples of the thermal flame thickness $\delta_{T}^0$ of a laminar freely-propagating flame under the respective conditions of the unburnt mixture:
\begin{itemize}[noitemsep]
    \item The domain height in $y$-direction is set to $L_y = 100 \delta_T^0$, ensuring that no length scales of the intrinsic instabilities are suppressed \cite{Berger2019}.
    \item The domain length in the streamwise direction is set to a sufficiently large value $L_x = 100 \delta_T^0$, to ensure that the global consumption speed of the freely-propagating flame remains unaffected by the domain length.
\end{itemize}
For all simulations the grid resolution is set to resolve $20$ points within the flame front.
The resulting computational grid for both the two-dimensional freely-propagating flame and the two-dimensional HOQ therefore consists of $2000$ cells in each direction, $n_x \times n_y = 2000 \times 2000$.

\begin{figure}[ht!]
    \begin{floatrow}
    \hfill
    \ffigbox[0.5\textwidth]{%
        \begin{tikzpicture}
            \begin{axis}[
                at={(-1.6cm,-0.4cm)},
                axis lines=none, 
                clip mode=individual, 
                width=5cm, 
                height=6.38cm, 
            ]
            
            \addplot[orange] table[col sep=comma, x index=0, y index=1] {data_for_figure2.csv};
            
            \end{axis}
        
            \draw[black, thick] (-2,0) rectangle (2,4);
            \node [] at (-1.6, -0.4) {\scriptsize $Y_{\mathrm{u}}$, $T_{\mathrm{u}}$};
            \draw[->] (-1.6, -0.2) -- (-1.3, 0.6);
            \node [] at (0.5, 2) {\scriptsize $Y_{\mathrm{b}}$, $T_{\mathrm{b}}$};
            \draw[thick](-2.2,0)--++(0.3,0);
            \draw[thick](-2.2,4)--++(0.3,0);
            \fill[pattern=north west lines] (-2.2,0) rectangle (-2, 4);
      
            \draw [-{Stealth}](-3.3,0) -- (-3.3,0.4);
            \draw [-{Stealth}](-3.3,0) -- (-2.9,0);
            \node [] at (-2.9, -0.15) {\scriptsize $x$};
            \node [] at (-3.5, 0.3) {\scriptsize $y$};
            \begin{scope}[>=Stealth]
                \draw [|<->|] (2.2, 0) -- (2.2, 4);
            \end{scope}
            \node [] at (2.5, 2) {\scriptsize $L_y$};
    
            \node[rotate=-90, color=gray] at (-1.8, 2) {\scriptsize wall};
            \node[rotate=90, color=gray] at (1.8, 2) {\scriptsize outlet};
            \node[color=gray] at (0.3, 4.2) {\scriptsize cyclic};
            \node[color=gray] at (0.3, -0.2) {\scriptsize cyclic};
        \end{tikzpicture}
    }
    {%
        \caption{ \footnotesize
            Schematic of the two-dimensional computational domain for the HOQ of thermodiffusively unstable flames, with the flame front mapped from the initial simulation of a freely-propagating flame.
            The domain dimensions are $L_x = 100 \delta_T^0$ and $L_y = 100 \delta_T^0$.
        }
        \label{fig:configuration_2D_HOQ}
    }
\hfill
\ffigbox[0.5\textwidth]{%
        \begin{tikzpicture}
            \draw[smooth, samples=800, domain=0:4, color=orange] plot({sin((\x)*20 r)*0.05+0.1}, \x);
            \draw[black, thick] (-2,0) rectangle (2,4);
            \node [] at (-1.1, 1.1) {\scriptsize $Y_{\mathrm{u}}$, $T_{\mathrm{u}}$};
            \node [] at (1.1, 1.1) {\scriptsize $Y_{\mathrm{b}}$, $T_{\mathrm{b}}$};
            \draw [-{Latex[round]}](-2.45,0.125) -- (-2,0.125);
            \draw [-{Latex[round]}](-2.45,0.375) -- (-2,0.375);
            \draw [-{Latex[round]}](-2.45,0.625) -- (-2,0.625);
            \draw [-{Latex[round]}](-2.45,0.875) -- (-2,0.875);
            \draw [-{Latex[round]}](-2.45,1.125) -- (-2,1.125);
            \draw [-{Latex[round]}](-2.45,1.375) -- (-2,1.375);
            \draw [-{Latex[round]}](-2.45,1.625) -- (-2,1.625);
            \draw [-{Latex[round]}](-2.45,1.875) -- (-2,1.875);
            \draw [-{Latex[round]}](-2.45,2.125) -- (-2,2.125);
            \draw [-{Latex[round]}](-2.45,2.375) -- (-2,2.375);
            \draw [-{Latex[round]}](-2.45,2.625) -- (-2,2.625);
            \draw [-{Latex[round]}](-2.45,2.875) -- (-2,2.875);
            \draw [-{Latex[round]}](-2.45,3.125) -- (-2,3.125);
            \draw [-{Latex[round]}](-2.45,3.375) -- (-2,3.375);
            \draw [-{Latex[round]}](-2.45,3.625) -- (-2,3.625);
            \draw [-{Latex[round]}](-2.45,3.875) -- (-2,3.875);
            \node [] at (-3.25, 2) {\scriptsize $u_{\mathrm{in}} \approx s_{\mathrm{c}}$};
            \draw [-{Stealth}](-3.3,0) -- (-3.3,0.4);
            \draw [-{Stealth}](-3.3,0) -- (-2.9,0);
            \node [] at (-2.9, -0.15) {\scriptsize $x$};
            \node [] at (-3.5, 0.3) {\scriptsize $y$};
            \begin{scope}[>=Stealth]
                \draw [>|-|<] (-0.13,0.1) -- (0.33,0.1);
                \draw [|<->|] (2.25, 0) -- (2.25, 4);
                \draw [|<->|] (-2, 4.15) -- (2, 4.15);
            \end{scope}
            \node [] at (0.1, -0.2) {\scriptsize $A_0$};
            \node [] at (2.55, 2) {\scriptsize $L_y$};
            \node [] at (0, 4.35) {\scriptsize $L_x$};
    
            \node[rotate=-90, color=gray] at (-1.8, 2) {\scriptsize inlet};
            \node[rotate=90, color=gray] at (1.8, 2) {\scriptsize outlet};
            \node[color=gray] at (1, 3.8) {\scriptsize cyclic};
            \node[color=gray] at (1, 0.2) {\scriptsize cyclic};
        \end{tikzpicture}
        }
        {%
        \caption{ \footnotesize
            Schematic of the two-dimensional computational domain used to generate thermodiffusively unstable flames in the non-linear regime.
            The initial flame front consists of weakly perturbed one-dimensional unstretched flames.
        }
        \label{fig:configuration_initial}
}
\end{floatrow}
\end{figure}

\subsection{Operating conditions}
\label{subsec:operating_conditions}

\begin{figure}
    \centering
    \includegraphics{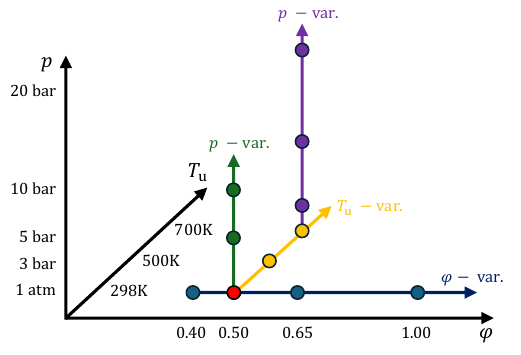}
    \caption{
        \footnotesize
        Visualization of the parametric variations of equivalence ratio $\varphi$, unburnt gas temperature $T_{\mathrm{u}}$, and pressure $p$.
        Reproduced from \cite{Berger2019}.   
    }
    \label{fig:parameter_space}
\end{figure}

Table \ref{tab:parametric_variations} provides a summary of the operating conditions employed in this study.
The operating conditions are similar to the cases investigated by Berger et al.~\cite{Berger2022, Berger2022a}.
For the reference case, the unburnt mixture is characterized by an equivalence ratio of $\varphi = 0.5 $, an unburnt temperature of $T_{\mathrm{u}} = \SI{298}{\kelvin}$, and a pressure of $p = \SI{1.01325}{\bar}$.
Parametric variations are conducted with respect to the equivalence ratio $\varphi$, the unburnt temperature $T_{\mathrm{u}}$, and the pressure $p$, while the other parameters are kept constant.
In addition to these three variations, a fourth variation of pressure at $\varphi = 0.5$ and $T_{\mathrm{u}} = \SI{700}{\kelvin}$ has been performed to include high-pressure, high-temperature conditions in the analysis, reflecting typical conditions for gas turbine applications \cite{Chiesa2005}.
Figure \ref{fig:parameter_space} shows the parametric space.

Table \ref{tab:parametric_variations} further presents characteristic properties of one-dimensional freely-propagating flames for the specified operating conditions:
the laminar burning velocity $s_{\mathrm{l}}$, the thermal flame thickness $\delta_T^0$, the characteristic flame time $\tau_T^0$, the flame power $q_{\mathrm{l}}$, the Zeldovich number $Ze$, the effective Lewis number ${Le}_{\mathrm{eff}}$ and the convection-diffusion Péclet number $Pe_{\mathrm{cd}}$.
The thermal flame thickness $\delta_T^0$ is defined as:
\begin{equation}
    \delta_T^0 = \frac{T_{\mathrm{b}} - T_{\mathrm{u}}}{\max\left(\mathrm{d}T/\mathrm{d}x\right)} \, ,
\end{equation}
with the adiabatic flame temperature $T_{\mathrm{b}}$.
The characteristic flame time $\tau_T^0$ is defined as the ratio between the thermal flame thickness and the laminar burning velocity $\tau_T^0 = {\delta_T^0}/{s_{\mathrm{l}}}$, and the flame power $q_{\mathrm{l}}$ is defined as:
\begin{equation}
    q_{\mathrm{l}} = \rho_{\mathrm{u}} s_{\mathrm{l}} \left( Y_{\ce{H2},\mathrm{u}} - Y_{\ce{H2},\mathrm{b}} \right) \left( c_{p,\mathrm{b}} T_{\mathrm{b}} -  c_{p,\mathrm{u}} T_{\mathrm{u}} \right ) \, ,
\end{equation}
where $\rho_{\mathrm{u}}$ is the density of the unburnt mixture, $Y_{\ce{H2},\mathrm{u}}$ and $Y_{\ce{H2},\mathrm{b}}$ denote the hydrogen mass fractions in the unburnt and burnt mixture and $c_{p,\mathrm{u}}$ and $c_{p,\mathrm{b}}$ refer to the specific heat capacities at constant pressure for the unburnt and burnt mixture.
The Zeldovich number $Ze$ is defined as:
\begin{equation}
    Ze = \frac{E}{R} \frac{(T_{\mathrm{b}} - T_{\mathrm{u}})}{T_{\mathrm{b}}^2} \, ,
\end{equation}
where $R$ is the universal gas constant and $E$ is the activation energy, that is approximated for a multi-step kinetic reaction based on the methodology outlined in \cite{Peters1987}:
\begin{equation}
    E = - 2 R \frac{\mathrm{d}(\rho_{\mathrm{u}} s_{\mathrm{l}})}{\mathrm{d}(1 / T_{\mathrm{b}})} \, .
    \label{eq:E_activation}
\end{equation}
According to \cite{Matalon2003, Sun1999}, the sensitivity coefficient in Eq. \ref{eq:E_activation} is evaluated from one-dimensional freely-propagating flame calculations by diluting the mixture with a small amount of nitrogen \ce{N2}, while the equivalence ratio, pressure, and unburnt temperature are kept constant.
The effective Lewis-number ${Le}_{\mathrm{eff}}$ for a lean mixture is calculated with the following formulation \cite{Matalon2007}:
\begin{equation}
    Le_{\mathrm{eff}} = \frac{Le_{\mathrm{O}} + Le_{\mathrm{F}} \mathcal{C}}{1 + \mathcal{C}} \, ,
\end{equation}
with the Lewis number of the fuel ${Le}_{\mathrm{F}}$ and oxidizer $Le_{\mathrm{O}}$ and the mixture strength $\mathcal{C}$ for a lean mixture:
\begin{equation}
    \mathcal{C} = 1 + Ze \left(\varphi^{-1} - 1\right) \, .
\end{equation}
In accordance with \cite{Berger2022, Berger2022a}, the Lewis numbers are taken from the burnt gas section of a one-dimensional freely-propagating flame.
In this case, the fuel Lewis number is defined as the Lewis number of hydrogen $Le_{\mathrm{F}} = Le_{\ce{H2}}$, and the oxidizer Lewis number is defined as the Lewis number of oxygen $Le_{\mathrm{O}} = Le_{\ce{O2}}$.
The convection-diffusion Péclet number is defined according to Rieth et al.~\cite{Rieth2023} as the ratio of convection to diffusion transport rates evaluated from one-dimensional freely-propagating flame calculations at corresponding conditions:
\begin{equation}
    Pe_{\mathrm{cd}} =\frac{\max{\left(\left| \mathrm{Conv.}_{\ce{H2}}      \right|\right)}}{\max{\left(\left| \mathrm{Diff.}_{\ce{H2}}    \right|\right)}} \, ,
\end{equation}
with the maximum absolute values of convective and diffusive fluxes $\mathrm{Conv.}_{\ce{H2}}$ and $\mathrm{Diff.}_{\ce{H2}}$.

\begin{table}[ht]
    \caption{
        \footnotesize
        Overview of the operating conditions (equivalence ratio $\varphi$, unburnt temperature $T_{\mathrm{u}}$, pressure $p$) of the simulations performed.
        For these conditions, the laminar burning velocity $s_{\mathrm{l}}$, the laminar flame thickness $\delta_T^0$, the characteristic flame time $\tau_T^0$, the flame power $q_{\mathrm{l}}$, the Zeldovich number $Ze$, the effective Lewis number ${Le}_{\mathrm{eff}}$ and the convection-diffusion Péclet number $Pe_{\mathrm{cd}}$ are also provided.
    }
    \small
    \begin{tabular}{lllllllllllll}
        \cline{1-11}
        Case & $\varphi$ & $p \,/\, \mathrm{bar}$ & $T_{\mathrm{u}} \,/\, \SI{}{\kelvin}$ & $s_{\mathrm{l}}\,/\, \SI{}{\frac{\meter}{\second}}$ & $\delta_T^0 \,/\, \SI{}{\micro\meter}$ & $\tau_T^0 \,/\, \SI{}{\milli \second}$ & $q_{\mathrm{l}} \,/\, \SI{}{\frac{\kilo W}{\meter^{2}}}$ & $Ze$ & ${Le}_{\mathrm{eff}}$ & $Pe_{\mathrm{cd}}$ & &  \\ \cline{1-11}
        \multicolumn{4}{l}{Reference:}                      & & & & & & & & \\
        {\color[HTML]{FE0000} Ref}                          & 0.5 & 1.01325 & 298 & 0.507 & 413 & 0.815 & 14.402 & 8.61 & 0.38 & 0.425 & &\\
        \multicolumn{4}{l}{Equivalence ratio variation:}    & & & & & & & & \\
        {\color[HTML]{00009B} phi0.4} (Part I \cite{PartI})  & 0.4 & 1.01325 & 298 & 0.224 & 610 & 2.21 & 4.218 & 11.14 & 0.34 & 0.289 & &\\
        {\color[HTML]{00009B} phi0.65}                       & 0.65 & 1.01325 & 298 & 1.013 & 346 & 0.342 & 45.938 & 6.61 & 0.46 & 0.600 & &\\
        {\color[HTML]{00009B} phi1.0}                        & 1.0 & 1.01325 & 298 & 2.042 & 344 & 0.169 & 176.537 & 6.19 & 0.75 & 0.855 & &\\
        \multicolumn{4}{l}{Unburnt temperature variation:}  & & & & & & & & \\                  
        {\color[HTML]{F8A102} Tu500}                        & 0.5 & 1.01325 & 500 & 1.883 & 412 & 0.219 & 32.150  & 5.69 & 0.41 & 0.764 & &\\
        {\color[HTML]{F8A102} Tu700}                        & 0.5 & 1.01325 & 700 & 5.235 & 474 & 0.091 & 63.733  & 3.31 & 0.46 & 1.379& &\\
        \multicolumn{4}{l}{\color[HTML]{6200C9} (= HighTuRef)} & & & & & & & & \\  
        \multicolumn{4}{l}{Pressure variation:}             & & & & & & & & \\              
        {\color[HTML]{036400} p05}                          & 0.5 & 5 & 298 & 0.253 & 112 & 0.441 & 35.525 & 12.34 & 0.36 & 0.247 & &\\
        {\color[HTML]{036400} p10}                          & 0.5 & 10 & 298 & 0.147 & 84 & 0.573 & 41.258 & 15.96 & 0.35 & 0.177 & &\\
        \multicolumn{4}{l}{Pressure variation at $T_{\mathrm{u}} = \SI{700}{\kelvin}$:} & & & & & & & & \\
        {\color[HTML]{6200C9} Tu700p03}                     & 0.5 & 3 & 700 & 4.052 & 120 & 0.030 & 147.153 & 4.80 & 0.43 & 0.975 & &\\
        {\color[HTML]{6200C9} Tu700p10}                     & 0.5 & 10 & 700 & 2.327 & 32 & 0.014 & 281.923 & 6.98 & 0.40 & 0.581 & &\\
        {\color[HTML]{6200C9} Tu700p20}                     & 0.5 & 20 & 700 & 1.468 & 18 & 0.012 & 355.896 & 8.52 & 0.38 & 0.412 & &\\ \cline{1-11}
    \end{tabular}
    \label{tab:parametric_variations}
\end{table}

\subsection{Numerical methods}
\label{subsec:numerical_methods}
The simulations in this work were performed using an in-house solver within the open-source CFD library OpenFOAM \cite{Weller1998}, which employs the Finite Volume Method (FVM) to solve the compressible reactive Navier-Stokes equations.
The chemical reactions are modeled by directly solving the chemical source terms using the detailed reaction mechanism of Li et al.~\cite{Li2004}, which includes $9$ species and $19$ reactions.
The governing equations for mass, momentum, species mass fractions, and enthalpy, along with the transport model and numerical schemes are identical to those presented in part I of this work \cite{PartI}.

\section{Parametric variations of the one-dimensional stable flames} \addvspace{10pt}
\label{sec:1D}
To demonstrate the effects of variations in equivalence ratio $\varphi$, unburnt temperature $T_{\mathrm{u}}$, and pressure $p$ on global flame properties and characteristic quantities of FWI, this section characterizes one-dimensional freely-propagating flames and one-dimensional HOQs for the parametric space presented in Tab. \ref{tab:parametric_variations}.
Furthermore, the quantities presented in this section serve as normalization values for the analyses of the two-dimensional HOQ in the subsequent section (Sec. \ref{sec:2DHOQ}).

The freely-propagating flames are briefly addressed first, followed by the HOQ.

\subsection{One-dimensional freely-propagating flames}
\label{subsec_1D}

\begin{figure}
    \centering
    \includegraphics[scale=0.9]{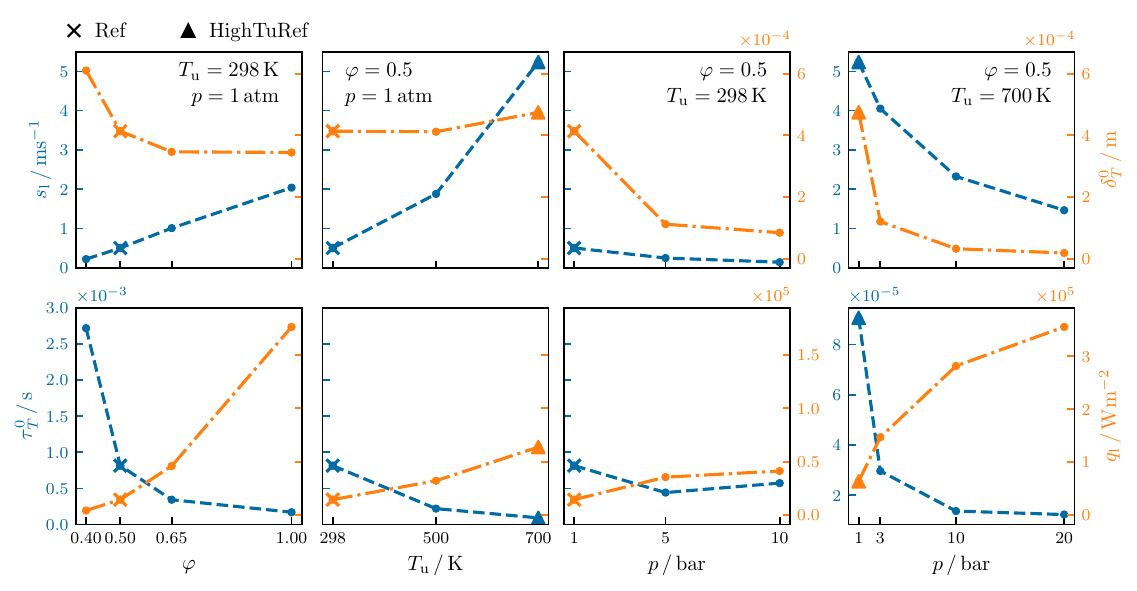}
    \caption{
    Top: Laminar flame speed $s_{\mathrm{l}}$ (left axis, blue) and thermal flame thickness $\delta_T^0$ (right axis, orange) for all parametric variations.
    Bottom:
    Characteristic flame time $\tau_T^0$ (left axis, blue) and flame power $q_{\mathrm{l}}$ (right axis, orange) for all parametric variations.
    The crosses indicate the reference case, `Ref`, the triangles highlight the reference case for the pressure variation at $T_{\mathrm{u}} = \SI{700}{K}$, `HighTuRef`.
    }
    \label{fig:fp_flame_paremeters}
\end{figure}

Figure \ref{fig:fp_flame_paremeters} shows the laminar flame speed $s_{\mathrm{l}}$, the thermal flame thickness $\delta_T^0$, the characteristic flame time $\tau_T^0$ and the flame power $q_{\mathrm{l}}$ for all operating conditions.
The following trends can be identified:
\begin{itemize}[noitemsep]
    \item \textbf{$\varphi$-variation}:
    With an increasing equivalence ratio (up to $\varphi = 1$) the flame speed increases almost linearly in accordance with \cite{Li2004} and the thermal flame thickness decreases.
    The characteristic flame time, calculated from the laminar flame speed and thermal flame thickness, consequently also decreases.
    The flame power strongly increases.
    \item \textbf{$T_{\mathrm{u}}$-variation}:
    The flame speed also increases with an increasing temperature of the unburnt mixture.
    The thermal flame thickness, however, remains nearly constant.
    Consequently, the flame time decreases.
    The flame power, in turn, also increases with an increasing $T_{\mathrm{u}}$.
    \item \textbf{$p$-variation}:
    With an increasing pressure, the laminar flame speed slightly decreases and the thermal flame thickness also decreases.
    As these two trends approximately balance each other, the characteristic flame time remains nearly constant (slight decrease followed by a small increase).
    The flame power also slightly increases with an increasing pressure.
    \item \textbf{$p$-variation at $T_{\mathrm{u}} = \SI{700}{\kelvin}$}:
    At $T_{\mathrm{u}} = \SI{700}{\kelvin}$ the trends for the laminar flame speed and thermal flame thickness are similar to those at $T_{\mathrm{u}} = \SI{700}{\kelvin}$, but especially the decrease in the laminar flame speed is significantly more pronounced.
    This leads to a notable decrease in the characteristic flame time.
    Also for this variation, the flame power increases with an increasing pressure.
\end{itemize}

\subsection{One-dimensional HOQ}
\label{subsec:1D_HOQ}

As these flames approach the wall and undergo quenching, the wall heat flux and quenching distance are key quantities, which are discussed in the following.

\begin{figure}
    \centering
    \includegraphics[scale=0.9]{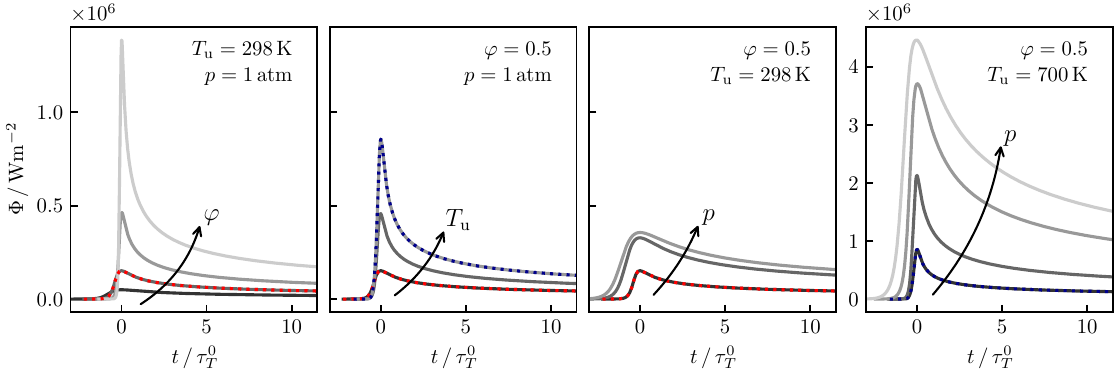}
    \caption{ \footnotesize
        Temporal wall heat flux evolution for the one-dimensional HOQs for the parametric variations (equivalence ratio $\varphi$, unburnt temperature $T_{\mathrm{u}}$ and pressure $p$).
        The time is relative to the quenching time and normalized with the characteristic flame time.
        The red dotted line highlights the reference case, `Ref`, the dark blue dotted line highlights the reference case for $T_{\mathrm{u}} = \SI{700}{K}$, `HighTuRef`.
    }
    \label{fig:whfs_over_time_1D}
\end{figure}

\begin{figure}
    \centering
    \includegraphics[scale=0.9]{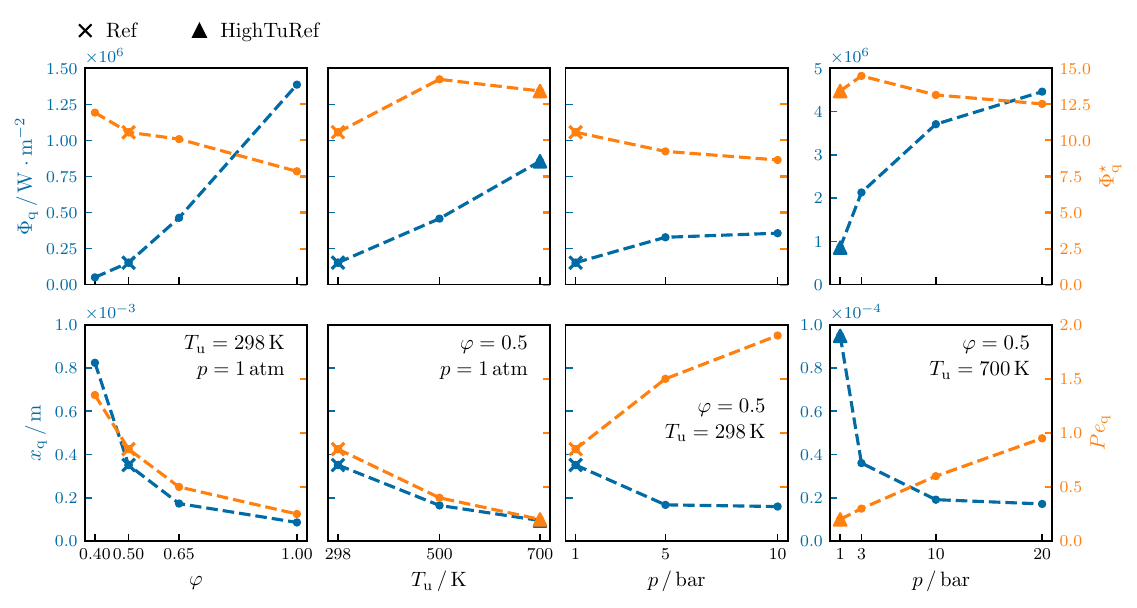}
    \caption{ \footnotesize
        Top: Quenching (maximum) wall heat flux (left axis, blue) and normalized quenching (maximum) wall heat flux (right axis, orange) of the one-dimensional HOQs for the variations of equivalence ratio $\varphi$, unburnt temperature $T_{\mathrm{u}}$ and pressure $p$ at $T_{\mathrm{u}} = \SI{298}{\kelvin}$ and $T_{\mathrm{u}} = \SI{700}{\kelvin}$.       
        Bottom: Quenching distance $x_{\mathrm{q}}$ (left axis, blue) and quenching Péclet number $Pe_{\mathrm{q}}$ (right axis, orange) of the one-dimensional HOQs for the parametric variations.
        The crosses indicate the reference case, `Ref`, the triangles highlight the reference case for $T_{\mathrm{u}} = \SI{700}{K}$, `HighTuRef`.
    }
    \label{fig:global_quantities_HOQ_1D}
\end{figure}

Figure \ref{fig:whfs_over_time_1D} illustrates the temporal profiles of wall heat flux during the one-dimensional HOQ process, highlighting the effects of varying equivalence ratio $\varphi$, unburnt gas temperature $T_{\mathrm{u}}$, and pressure $p$ on the quenching dynamics.
Additionally, Figure \ref{fig:global_quantities_HOQ_1D} shows the quenching wall heat flux $\Phi_{\mathrm{q}}$, the normalized quenching wall heat flux $\Phi_{\mathrm{q}}^{\star}$ (top), the quenching distance $x_{\mathrm{q}}$ and the quenching Péclet number $Pe_{\mathrm{q}}$ (bottom), to quantify trends in global parameters of the quenching process for the parametric variations of the one-dimensional HOQ.
The quenching wall heat flux $\Phi_{\mathrm{q}}$ is defined as the temporal maximum of the wall heat flux ($\Phi_{\mathrm{q}} = \max_t(\Phi(t)$), the quenching distance $x_{\mathrm{q}}$ is defined as the temporal minimum of the wall distance of the flame front.
The flame front is defined by the location of the maximum absolute value of the \ce{H2} source term $\dot{\omega}_{\ce{H2}}$ in this study.
The normalized wall heat flux $\Phi^{\star}$ is normalized by the flame power $q_{\mathrm{l}}$ of the freely-propagating flame at the respective operating condition and the quenching Péclet number $Pe_{\mathrm{q}}$ is defined as the quenching distance $x_{\mathrm{q}}$ normalized by the thermal flame thickness $\delta_T^0$ of the laminar flame at identical operating conditions.

The following trends are observed for the temporal quenching processes and global quenching quantities:
\begin{itemize} [noitemsep]
    \item \textbf{$\varphi$-variation}:
    With increasing equivalence ratio $\varphi$, the flame power $q_{\mathrm{l}}$ increases and the thermal flame thickness $\delta_T^0$ decreases (see Fig. \ref{fig:fp_flame_paremeters}), which results in a steeper temperature gradient at the wall during the quenching process, thereby increasing the wall heat flux (see Fig. \ref{fig:whfs_over_time_1D}) and its maximum value $\Phi_\mathrm{q}$ (see Fig. \ref{fig:global_quantities_HOQ_1D}).
    The normalized quenching wall heat flux $\Phi_{\mathrm{q}}^{\star}$ slightly decreases.
    The decreasing quenching distance $x_{\mathrm{q}}$ is consistent with the increasing quenching wall heat flux $\Phi_{\mathrm{q}}$.
    The quenching Péclet number $Pe_{\mathrm{q}}$ also decreases with increasing equivalence ratio $\varphi$, which can be attributed to the minimal variations in thermal flame thickness $\delta_T^0$ relative to the quenching distance $x_{\mathrm{q}}$ (see Fig. \ref{fig:fp_flame_paremeters}).
    \item \textbf{$T_{\mathrm{u}}$-variation}:
    With increasing unburnt gas temperature $T_{\mathrm{u}}$, the thermal flame thickness $\delta_T^0$ is nearly constant, but the adiabatic flame temperature and the flame power $q_{\mathrm{l}}$ increase, also leading to a higher temperature gradient at the wall and therefore a higher wall heat flux $\Phi(t/\tau_T^0)$ (see Fig. \ref{fig:whfs_over_time_1D}) and quenching wall heat flux $\Phi_{\mathrm{q}}$ (see Fig. \ref{fig:global_quantities_HOQ_1D}).
    There is no clear trend for the normalized quenching wall heat flux $\Phi_{\mathrm{q}}^{\star}$, whereas the trends for the quenching distance $x_{\mathrm{q}}$ and the quenching Péclet number $Pe_{\mathrm{q}}$ are similar to those observed in the $\varphi$-variation.
    \item \textbf{$p$-variation}:
    Also with an increasing pressure $p$, the wall heat flux increases, in general (see Fig. \ref{fig:whfs_over_time_1D}) and its peak value $\Phi_{\mathrm{q}}$ (see Fig. \ref{fig:global_quantities_HOQ_1D}), which can be attributed to the decreasing thermal flame thickness $\delta_T^0$ and the increasing flame power $q_{\mathrm{l}}$.
    Consequently, the quenching distance $x_{\mathrm{q}}$ decreases.
    The normalized quenching wall heat flux $\Phi_{\mathrm{q}}^{\star}$ slightly decreases, whereas the quenching Péclet number $Pe_{\mathrm{q}}$ slightly increases due to the more pronounced decrease in thermal flame thickness compared to the quenching distance $x_{\mathrm{q}}$ with an increasing pressure.
    \item \textbf{$p$-variation at $T_{\mathrm{u}} = \SI{700}{\kelvin}$}:
    The same trends observed for the $p$-variation at $T_{\mathrm{u}} = \SI{298}{\kelvin}$ also apply here; however, the elevated unburnt gas temperature of $T_{\mathrm{u}} = \SI{700}{\kelvin}$ leads to even higher values of the wall heat flux $\Phi(t)$ (see Fig. \ref{fig:whfs_over_time_1D}) and quenching wall heat flux $\Phi_{\mathrm{q}}$ and even lower values of the quenching distance $x_{\mathrm{q}}$ (see Fig. \ref{fig:global_quantities_HOQ_1D}) (note the adjusted scale in this plots). 
\end{itemize}

In summary, the quenching distance $x_{\mathrm{q}}$ decreases with increasing equivalence ratio $\varphi$, unburnt temperature $T_{\mathrm{u}}$, and pressure $p$, and the quenching wall heat flux $\Phi_{\mathrm{q}}$ increases accordingly.
These trends particularly provide references for the parametric variations in the two-dimensional HOQ.

\section{Parametric variations of the two-dimensional unstable flames} \addvspace{10pt}
\label{sec:2DHOQ}

After the discussion of the one-dimensional flames, the effects of the instabilities on the FWI are analyzed based on the two-dimensional HOQ configuration (Sec. \ref{sec:NumericalSetup}).
First, key characteristics of unstable flames prior to quenching are discussed, focusing on the effects of the parametric variations.
Subsequently, the influence of instabilities on FWI is assessed across the varying operating conditions.

\subsection{Characterization of the unstable hydrogen/air flames prior to quenching}

\begin{figure}
    \centering
    \includegraphics[scale=0.8]{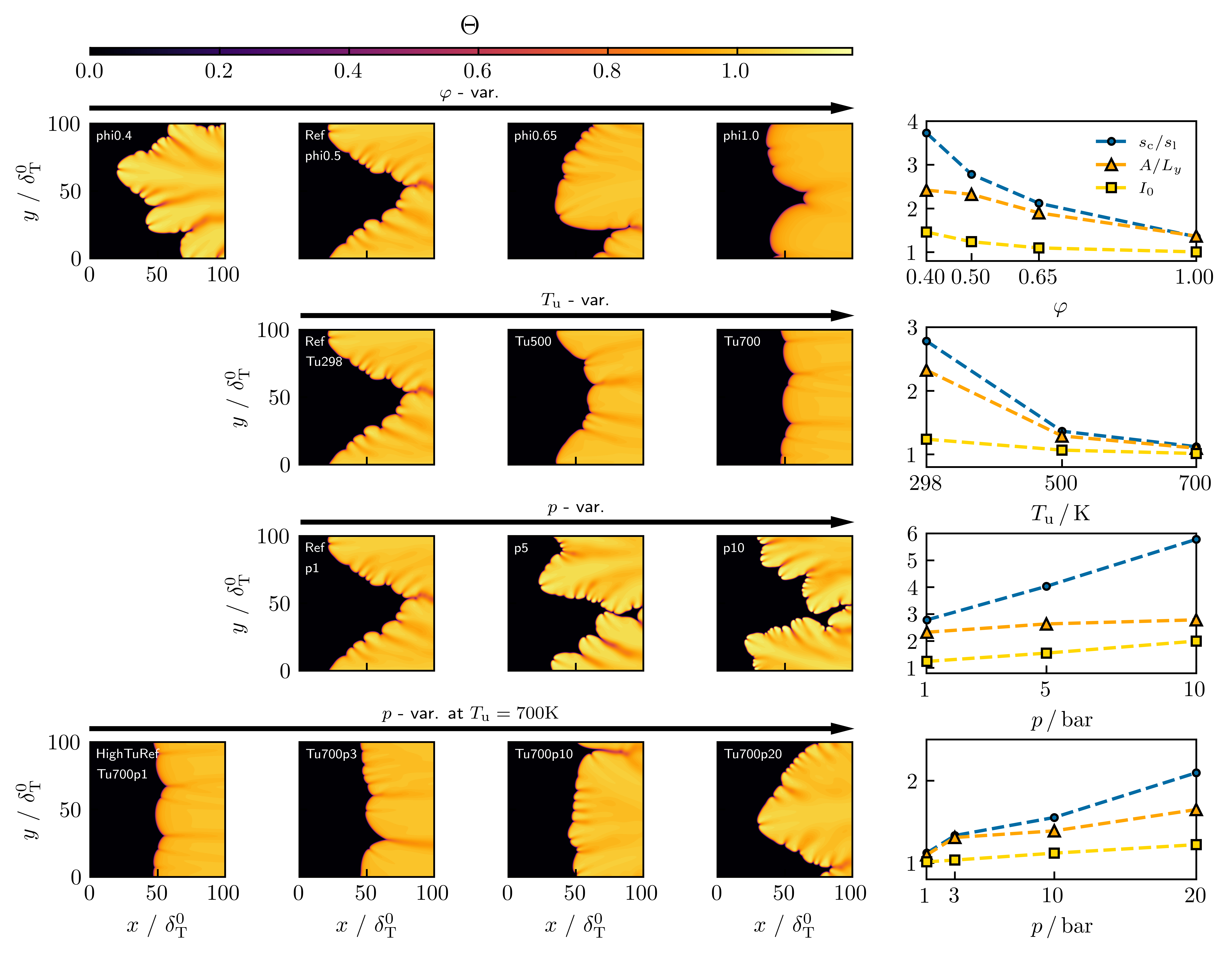}
    \caption{ \footnotesize
        Left: Profiles of the normalized temperature $\Theta$ over for the two-dimensional freely-propagating unstable flames, which serve as the initial condition for the two-dimensional HQO.
        Right: Normalized consumption speed $s_{\mathrm{c}} / s_{\mathrm{l}}$, the normalized flame surface area $A / L_y$ and the reactivity factor $I_0$ for the respective parametric variation (equivalence ratio $\varphi$, temperature of the unburnt mixture $T_{\mathrm{u}}$, pressure $p$ $T_{\mathrm{u}} = \SI{298}{\kelvin}$ and pressure $p$ at $T_{\mathrm{u}} = \SI{700}{\kelvin}$).
    }
    \label{fig:overview_unstable_flames_sl_I0}
\end{figure}

To asses the effects of variations in equivalence ratio $\varphi$, unburnt gas temperature $T_{\mathrm{u}}$ and pressure $p$ on the unstable flames prior to quenching, the flame fronts and key quantities of thermodiffusively unstable flames are introduced to characterize the varying operating conditions.
Since the operating conditions are identical to those analyzed in detail by Berger et al.~\cite{Berger2022a}, only the key findings are highlighted to provide a baseline for investigating the HOQ process in unstable flames.
Fig. \ref{fig:overview_unstable_flames_sl_I0} (left) shows the normalized temperature $\Theta$ over the entire numerical domain for a single snapshot for the two-dimensional freely-propagating flame for each operating condition.
The normalized temperature is defined as $\Theta = {\left(T - T_{\mathrm{u}}\right)}/{\left(T_{\mathrm{b}} - T_{\mathrm{u}}\right)}$, with the adiabatic flame temperature $T_{\mathrm{ad}}$ for the respective operating condition.
Figure \ref{fig:overview_unstable_flames_sl_I0} (right) additionally shows the normalized consumption speed $s_{\mathrm{c}}/s_{\mathrm{l}}$, the normalized flame surface $A / L_{y}$ and the reactivity factor $I_0$ for each parametric variation.
The normalized consumption speed is normalized by the laminar flame speed $s_{\mathrm{l}}$ of a one-dimensional freely-propagating flame at the identical operating condition (see Fig. \ref{fig:fp_flame_paremeters} and Tab. \ref{tab:parametric_variations}).
In accordance with Howarth and Aspden \cite{Howarth2022}, the flame surface $A$ in two dimensions is defined as the length of the isoline of $\mathrm{PV} = 1 - \frac{Y_{\ce{H2}}}{Y_{\ce{H2}\mathrm{,u}}} = 0.9$, with the \ce{H2} mixture fraction in the unburnt mixture $Y_{\ce{H2}\mathrm{,u}}$.
It is normalized by the area of the inlet, which collapses to $L_{y}$ in two dimensions.
The reactivity factor $I_0$ is defined as:
\begin{equation}
    I_0 = \frac{s_{\mathrm{c}}}{s_{\mathrm{l}}}\bigg / \frac{A}{L_{y}} \, .
\end{equation}
While the normalized consumption speed $s_{\mathrm{c}}/s_{\mathrm{l}}$ includes the increase in flame surface area $A/L_y$, the reactivity factor $I_0$ quantifies variations of the local reactivity and flame structure, i.e. the deviations of the local thermochemical states from a stable, one-dimensional freely-propagating flamelet \cite{PartI, Berger2023}. 

The following trends are highlighted for the different parametric variations, given their importance to this study:
\begin{itemize}[noitemsep]
    \item \textbf{$\varphi$-variation}:
    With an increasing equivalence ratio $\varphi$, the flame front corrugation and the magnitude of the superadiabatic temperatures ($\Theta>1)$ decrease, indicating the decreasing intensity of thermodiffusive instabilities.
    This is further confirmed by the profiles of $s_{\mathrm{c}}/s_{\mathrm{l}}$, $A/L_y$ and $I_0$.
    For low values of the equivalence ratio, the increased consumption speed $s_{\mathrm{c}} / s_{\mathrm{l}}$ is attributed to both the enhancement of the flame surface area $A/L_y$ due to wrinkling and an increase in reactivity $I_0$, indicating the high intensity of thermodiffusive instabilities.
    As the equivalence ratio increases, $s_{\mathrm{c}}/s_{\mathrm{l}}$, $A/L_y$, and $I_0$ decrease, quantitatively confirming a reduction in thermodiffusive instability intensity.
    At $\varphi = 1$, the increased normalized global consumption speed $s_{\mathrm{c}} / s_{\mathrm{l}} > 1$ is primarily due to increased flame surface area $A/L_y$ driven by hydrodynamic instabilities, with the reactivity factor $I_0 = 1$ indicating minimal thermodiffusive effects.
    \item \textbf{$T_{\mathrm{u}}$-variation}:
    The trends observed for increasing unburnt gas temperature $T_{\mathrm{u}}$ are similar to those for increasing equivalence ratio $\varphi$, as the intensity of the instabilities, quantified by $I_0$, also decreases.
    At $T_{\mathrm{u}} = \SI{500}{\kelvin}$ the wrinkling of the flame front is significantly decreased and at $T_{\mathrm{u}} = \SI{700}{\kelvin}$ the flame front is characterised exclusively by hydrodynamic instabilities.
    Accordingly, $I_0$ approaches a value of $I_0 = 1$ with increasing unburnt gas temperature $T_{\mathrm{u}}$.
    \item \textbf{$p$-variation}:
    As the pressure $p$ increases, the corrugation of the flame front intensifies significantly.
    The increasing intensity of thermodiffusive instabilities is also evident from the superadiabatic temperatures ($\Theta > 1$) and the small scale structures, which increase with increasing pressure $p$.
    Consequently, in addition to the global consumption speed $s_{\mathrm{c}} / s_{\mathrm{l}}$ and the increase in the flame surface area $A / L_{y}$, the reactivity factor $I_0$ also increases up to a value of around $2$ at $p = \SI{20}{\bar}$.
    \item \textbf{$p$-variation at $T_{\mathrm{u}} = \SI{700}{\kelvin}$}:
    The trends for a pressure variation at $T_{\mathrm{u}} = \SI{700}{\kelvin}$ are similar to those at $T_{\mathrm{u}} = \SI{298}{\kelvin}$.
    The difference is that in the reference case `HighTuRef` at $p = \SI{1}{\bar}$ thermodiffusive instabilities are not pronounced and hydrodynamic instabilities are predominant.
    Since the increase in pressure and the increase in temperature have an opposing effect on thermodiffusive instabilities, the wrinkling of the flame front and also the global consumption speed $s_{\mathrm{c}}/s_{\mathrm{l}}$, the enlargement of the flame surface $A/L_y$ and the reactivity factor $I_0$ increase significantly with increasing pressure $p$, but do not reach the same values as for $T_{\mathrm{u}} = \SI{298}{\kelvin}$.
\end{itemize}
These observations are consistent with previous studies by Berger et al.~\cite{Berger2019, Berger2023} and highlight the significant impact of equivalence ratio, temperature, and pressure on intrinsic instabilities, which is crucial for the subsequent analysis of the quenching process.

\subsection{Analysis of the quenching process}
This section discusses the influence of intrinsic instabilities on the quenching process across all parametric variations, which correspond to flames with varying intensities of thermodiffusive instabilities as shown in Fig. \ref{fig:overview_unstable_flames_sl_I0}.
First, the quenching process is illustrated for four representative operating conditions, each showing a different intensity of the instabilities.
Following this, the effect of the intrinsic instabilities on the global parameters of the quenching process is analyzed for all operating conditions, with a focus on the varying intensity of the instabilities.

\begin{figure}
    \centering
    \includegraphics{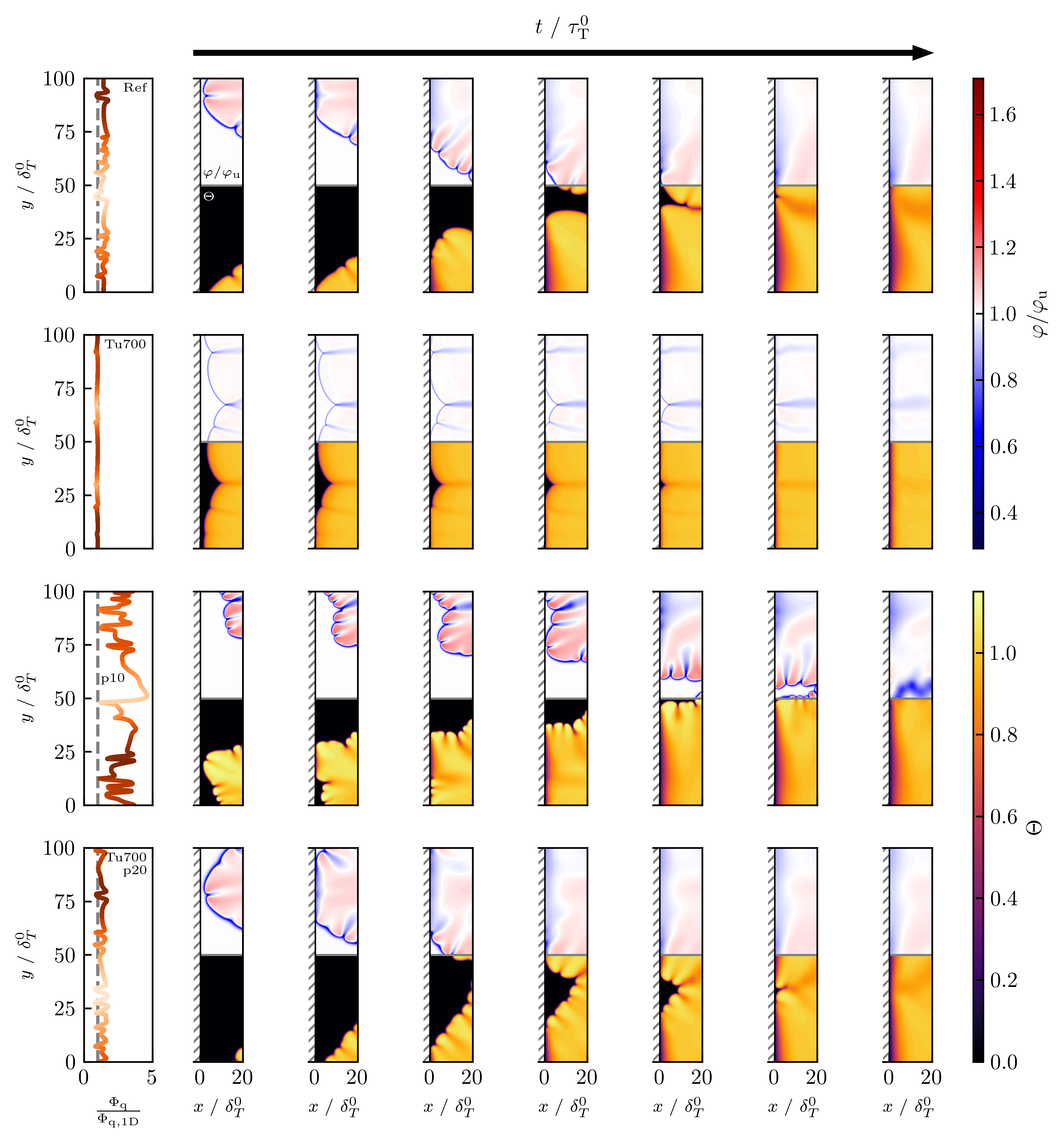}
    \caption{ \footnotesize
        Left: Quenching wall heat flux $\Phi_{\mathrm{q}}$ along $y$ normalized by the quenching wall heat flux $\Phi_{\mathrm{1D}}$ of the one-dimensional HOQ at the respective operating condition.
        The color scale represents the quenching time step for the respective wall position, with darker colors indicating earlier time steps.
        Right: Normalized local equivalence ratio $\varphi / \varphi_{\mathrm{u}}$ ($y/\delta_T^0 > 50$) and normalized temperature $\Theta$ ($y/\delta_T^0 < 50$) over the numerical domain for different time instances of the quenching process for the four cases `Ref`, `Tu700`, `p10` and `Tu700p20`. 
        The location of the wall is indicated by grey dashed lines.
        The direction of increasing time is indicated by the black arrow over the profiles.
        Animations of the time evolution for each case can be found in the Supplementary Material.    
    }
    \label{fig:HOQ_over_domain}
\end{figure}
Figure \ref{fig:HOQ_over_domain} shows consecutive snapshots of the quenching process and the normalized quenching wall heat flux $\Phi_{\mathrm{q}} / \Phi_{\mathrm{q,1D}}$.
Four representative cases are shown: `Ref` shows a moderate level of instabilities and serves as the reference for the parametric variations.
`Tu700` is predominantly characterized by hydrodynamic instabilities.
`p10` shows high intensity of thermodiffusive instabilities.
`Tu700p20` is an example of operation conditions found in technical applications, where the reduction of instability intensity caused by the increase in unburnt gas temperature counteracts the increased instability intensity caused by the elevated pressure.

A detailed analysis of the quenching process for `phi0.4` was carried out in part I of this work \cite{PartI}.
It was shown that thermodiffusive instabilities lead to a complex quenching process, with the structures of the instabilities significantly influencing the local wall heat flux.
The following quenching processes can be observed for the presented cases:
\begin{itemize} [noitemsep]
    \item `Ref`:
    The `Ref` case is similar to the one analyzed in part I, exhibiting a marginally lower level of thermodiffusive instabilities due to the higher equivalence ratio ($\varphi = 0.5$ compared to $\varphi = 0.4$) and the resulting increase in the effective Lewis number $Le_{\mathrm{eff}}$.
    Still, the reactivity factor $I_0$ is significantly larger than $1$.
    First, a flame finger approaches the wall and quenches in HOQ manner.
    From this location, the flame front propagates downwards and upwards and quenches in side-wall quenching (SWQ) \cite{Dreizler2015} manner.
    Finally, different areas of the flame front merge, resulting in a locally very lean mixture that quenches in HOQ manner.
    The structure of the flame fingers is reflected in the normalized quenching wall heat flux; the wall heat flux is significantly greater in the positively curved areas than in the negatively curved area at the flame tip.
    In general, the average quenching wall heat flux is significantly larger than in the corresponding one-dimensional HOQ under the same operating conditions.
    \item `Tu700`:
    For `Tu700`, the local mixture variation is significantly lower than for `Ref` due to the absence of thermodiffusive instabilities ($I_0 \approx 1$).
    The entire flame front approaches the wall with an almost uniform local flame speed and quenches uniformly in a HOQ-like manner, albeit at slightly varying angles.
    Although certain areas reach the wall first, there is no propagation along the wall and no quenching in a SWQ manner; instead, the trailing areas quench in an HOQ manner with a time delay.
    The normalized quenching wall heat flux is almost constant along the wall and corresponds to the value from the one-dimensional HOQ under the same operating conditions.
    \item `p10`:
    This case is characterized by a high level of thermodiffusive instabilities (high reactivity factor $I_0$).
    First, a flame finger approaches the wall and quenches in HOQ manner (lower part of the domain), similar to case `Ref`.
    Subsequently, the flame propagates locally both upward and downward from the initial quenching location and quenches in SWQ manner.
    Simultaneously, a second region of the flame front interacts with the wall and also quenches in HOQ manner (upper part of the domain), followed by a second local flame front propagating along the wall and quenching in a SWQ manner.
    During this propagation along the wall, very high quenching wall heat fluxes are observed, accompanied by high local flame curvature and variations of the local equivalence ratio $\varphi / \varphi_{\mathrm{u}}$.
    Finally, the two flame fronts interact and merge, resulting in a locally very lean mixture and a comparatively small quenching wall heat flux.
    Due to the highly pronounced wrinkling of the flame front and small-scale structures, the variation in the quenching wall heat flux is significantly more pronounced compared to the reference case `Ref`.
    The quenching wall heat flux reaches values up to five times higher than those observed in one-dimensional HOQ under identical operating conditions.
    \item `Tu700p10`:
    The case is very similar to the `Ref` case, both in the level of thermodiffusive instabilities and the temporal evolution of the quenching process and the normalized quenching wall heat flux $\Phi_{\mathrm{q}}/\Phi_{\mathrm{q,1D}}$.
\end{itemize}
This analysis of the quenching process further supports the findings from part I \cite{PartI}, indicating that the intensity of thermodiffusive instabilities has a significant impact on the quenching process, especially on the relative increase in the wall heat flux compared to a corresponding one-dimensional head-on quenching of a stable flame.

\begin{figure}
    \centering
    \includegraphics{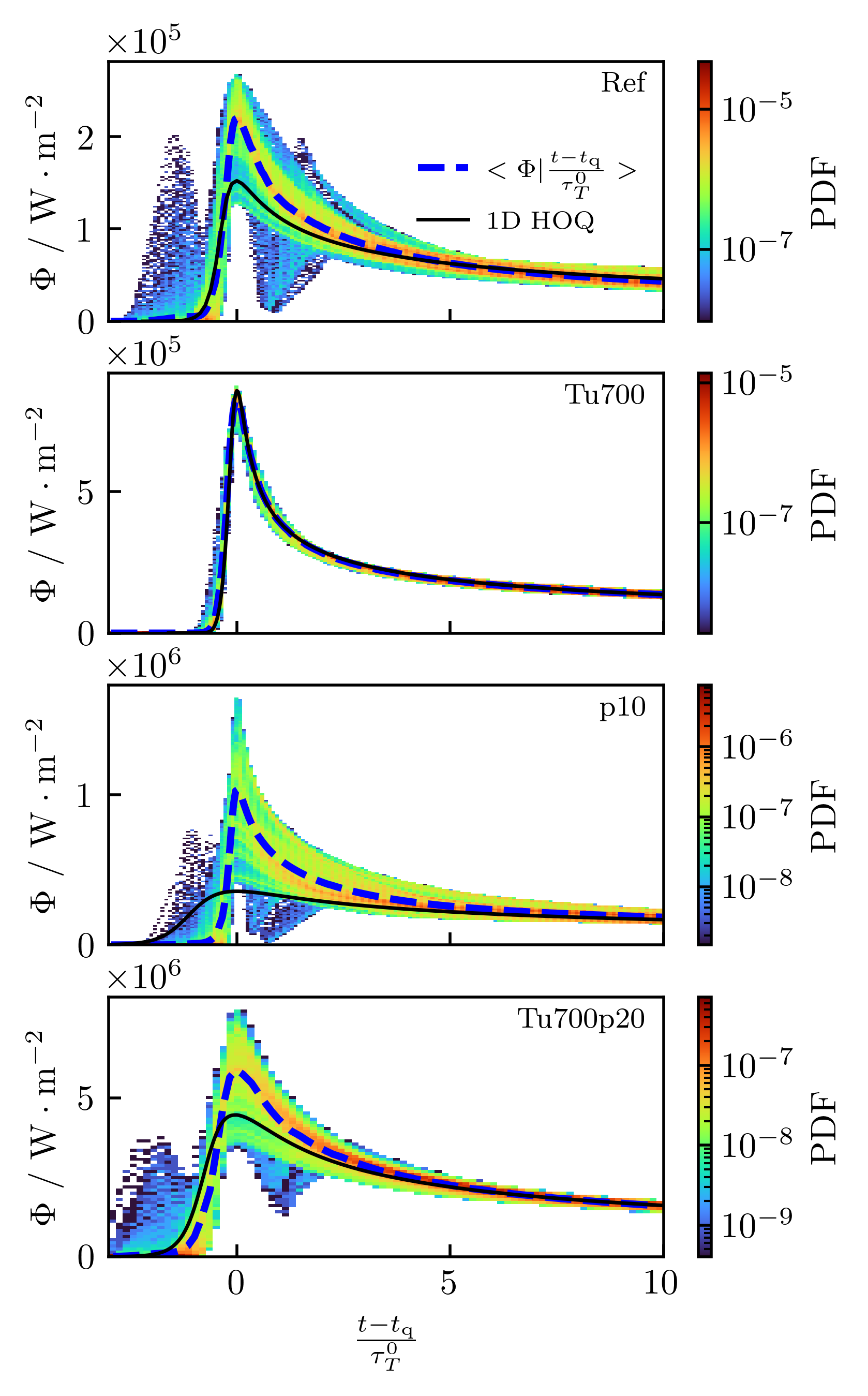}
    \caption{\footnotesize
        PDFs of the wall heat flux $\Phi$ over a relative normalized time $\left(t-t_{\mathrm{q}}\right)/\tau_T^0$ (relative to the quenching time $t_{\mathrm{q}}$ and normalized by the characteristic flame time $\tau_T^0$ of a one-dimensional freely-propagating flame at the respective operating condition) for the cases shows in Fig. \ref{fig:HOQ_over_domain} (`Ref`, `Tu700`, `p10` and `Tu700p20`).
        The dashed blue line indicates the wall heat flux $\Phi$ conditioned on the normalized relative time $\left(t-t_{\mathrm{q}}\right) / \tau_T^0$.
        The black line corresponds to a one-dimensional HOQ at the respective operating condition.
    }
    \label{fig:quenching_whf_local_examples}
\end{figure}
To further illustrate that the intensity of thermodiffusive instabilities impacts not only the quenching wall heat flux $\Phi_{\mathrm{q}}$ but also the temporal progression of the wall heat flux, Fig. \ref{fig:quenching_whf_local_examples} shows the evolution of the wall heat flux $\Phi$ over a normalized relative time, ${(t-t_{\mathrm{q}})}/{\tau_{T}^0}$.
The joint PDF corresponds to all the grid cells along the wall $y$ for the two-dimensional HOQs.
The following local wall heat flux characteristics can be identified for the presented cases:
\begin{itemize} [noitemsep]
    \item `Ref`:
    The progression of the quenching wall heat flux qualitatively matches the profile of `phi0.4`, which was thoroughly discussed in part I \cite{PartI} of this study.
    However, the scatter around the mean profile is reduced, consistent with the lower intensity of thermodiffusive instabilities (see Fig. \ref{fig:overview_unstable_flames_sl_I0} (right)).
    Double quenching events, which lead to the regions with low PDF values and are characteristic of thermodiffusive instabilities (see part I), also occur for this case.
    The PDF further illustrates that thermodiffusive instabilities lead to higher quenching wall heat fluxes not only on average ($<\Phi|(t-t_{\mathrm{q}})/\tau_T^0>$) but at nearly all locations, compared to a stable flame (one-dimensional reference).
    After the quenching event, the mean wall heat flux profile gradually converges with that of the stable flame, reaching alignment after several flame times.    
    \item `Tu700`:
    In the case of `Tu700`, which is predominantly influenced by hydrodynamic instabilities, the entire PDF of the quenching process collapses to the average profile $<\Phi|(t-t_{\mathrm{q}})/\tau_T^0>$, which closely aligns with that of the stable flame (one-dimensional reference).
    This further confirms that hydrodynamic instabilities have a minor influence on the local quenching wall heat flux profile.
    \item `p10`:
    For the case `p10`, the maximum values of the wall heat flux increase noticeably, consistent with a higher intensity of thermodiffusive instabilities (see Fig. \ref{fig:overview_unstable_flames_sl_I0}, right).
    Double quenching events (see part I \cite{PartI}) are also observed in this case.
    From the quenching point onward, the PDF values are almost entirely above those of the stable HOQ flame.
    The difference between the mean profile of the unstable HOQ flame and the stable HOQ flame (one-dimensional reference) is even more pronounced here than in the `Ref` case. 
    \item `Tu700p10`:
    For `Tu700p10`, the maximum values of the wall heat flux decrease significantly compared to `p10`, closely aligning with the reduction in the intensity of the thermodiffusive instabilities (see Fig. \ref{fig:overview_unstable_flames_sl_I0} (right) and Fig. \ref{fig:HOQ_over_domain}).
    The mean wall heat flux profile shows much closer agreement with the stable one-dimensional HOQ profile compared to case `p10`.
    Overall, both the PDF and the mean wall heat flux profile show strong qualitative agreement with the `Ref` case.    
\end{itemize}
In conclusion, the intensity of thermodiffusive instabilities also clearly influences the local temporal progression of the quenching process.

After the discussion of local quenching phenomena, the impact of instabilities on global quenching characteristics is evaluated and quantified, with a specific focus on the quenching wall heat flux $\Phi_{\mathrm{q}}$ and the quenching distance $x_{\mathrm{q}}$.

\begin{figure}
    \centering
    \includegraphics[scale=0.8]{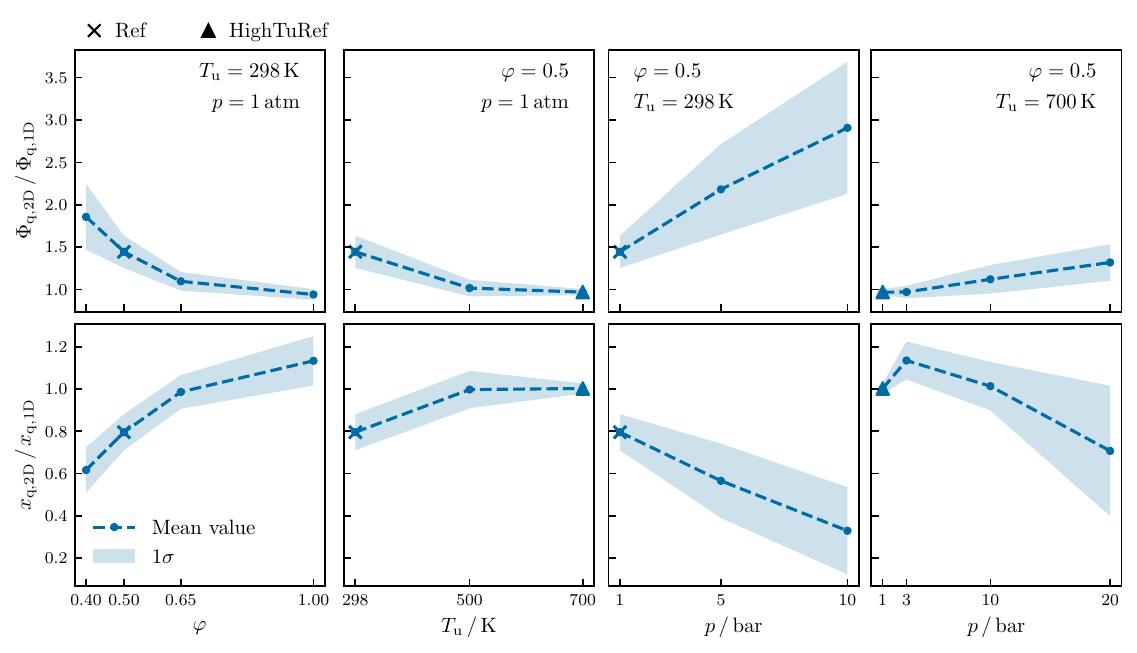}
    \caption{ \footnotesize
        Top: Mean of the quenching wall heat flux $\overline{\Phi}_{\mathrm{q,2D}}$ $\pm 1 \sigma$ (standard deviation) relative to the quenching wall heat flux for the one-dimensional HOQ $\Phi_{\mathrm{q,1D}}$ for the respective operating condition.
        Bottom: Mean quenching distance $\overline{x}_{\mathrm{q,2D}}$ $\pm 1 \sigma$ (standard deviation) relative to the quenching distance of the one-dimensional HOQ $x_{\mathrm{q,1D}}$.
        The regions shaded in light blue correspond to one standard deviation.
    }
    \label{fig:whf_and_quenching_distance_2D_to_1D}
\end{figure}

Figure \ref{fig:whf_and_quenching_distance_2D_to_1D} shows the relative quenching wall heat flux and quenching distance (mean value $\pm 1$ standard deviation) from the two-dimensional HQOs compared to the respective one-dimensional HQO.
The impact of the thermodiffusive instabilities on the quenching characteristics is reflected in the differences observed between the one- and two-dimensional configurations: while instabilities are inherently suppressed in the one-dimensional HOQ, they can develop in the two-dimensional HQO.
Furthermore, the absolute quenching wall heat fluxes and quenching distances are shown in Fig. \ref{fig:overview_whf_qd_norm_and_nonnorm} in the \ref{sec:app_add_reps} for the one-dimensional and two-dimensional HOQs to highlight the influence of instabilities on the quenching process in comparison to the effects of variations in the operating conditions.
The following trends are identified:
\begin{itemize} [noitemsep]
    \item \textbf{$\varphi$-variation}:
    For lean operating conditions, the quenching process is clearly influenced by the thermodiffusive instabilities, leading to higher relative quenching wall heat fluxes.
    With an increasing equivalence ratio $\varphi$, the mean relative quenching wall heat flux and the region of $\pm 1$ standard deviation decrease.
    Consistently, the quenching distance increases.  
    While the mean quenching wall heat flux for $\varphi = 0.4$ in the two-dimensional case $\overline{\Phi}_{\mathrm{q,2D}}$ is approximately twice as high as in the one-dimensional case ${\Phi_{\mathrm{q,1D}}}$ and for the reference case (`Ref`, $\varphi = 0.5$) $\overline{\Phi}_{\mathrm{q,2D}} / \Phi_{\mathrm{q,1D}} \approx 1.5 $, there is no relative increase for $\varphi = 1$.
    Notably, the relative quenching wall heat flux decreases and the quenching distance increases as the intensity of thermodiffusive instabilities, quantified by the reactivity factor $I_0$ (see Fig. \ref{fig:overview_unstable_flames_sl_I0}, right), decreases.
    \item \textbf{$T_{\mathrm{u}}$-variation}:
    A similar trend is observed for an increasing $T_{\mathrm{u}}$.
    As $T_{\mathrm{u}}$ is increased, the mean relative quenching wall heat flux $\overline{\Phi}_{\mathrm{q,2D}} / \Phi_{\mathrm{q,1D}}$ decreases, similar to the intensity of thermodiffusive instabilities, quantified by the reactivity factor $I_0$ (see Fig. \ref{fig:overview_unstable_flames_sl_I0} (right)).
    For $T_{\mathrm{u}} = \SI{700}{\kelvin}$ there is also no relative increase in the mean quenching wall heat flux ($\overline{\Phi}_{\mathrm{q,2D}} / \Phi_{\mathrm{q,1D}} \approx 1$) and no relative decrease in the quenching distance ($\overline{x}_{\mathrm{q,2D}}/x_{\mathrm{q,1D}} \approx 1$).
    Consequently, there is only marginal influence of hydrodynamic instabilities, which are predominantly present at these operating points, on the quenching process, as indicated in Fig. \ref{fig:HOQ_over_domain}.
    \item \textbf{$p$-variation}:
    However with an increasing pressure $p$, the mean values and standard deviation of the relative quenching wall heat flux ${\Phi_{\mathrm{q,2D}}} / \Phi_{\mathrm{q,1D}}$ significantly increase, reaching values $3$ times higher than in the one-dimensional case for $p = \SI{10}{\bar}$.
    Consequently, the relative quenching distance decreases with increasing pressure.
    This trend is also consistent with the increasing intensity of thermodiffusive instabilities with increasing pressure, quantified by the increasing values of the reactivity factor $I_0$ in Fig. \ref{fig:overview_unstable_flames_sl_I0} (right).
    \item \textbf{$p$-variation at $T_{\mathrm{u}} = \SI{700}{\kelvin}$}:
    The trends for increasing pressure $p$ at $T_{\mathrm{u}} = \SI{700}{\kelvin}$ are similar to those observed at $T_{\mathrm{u}} = \SI{298}{\kelvin}$.
    As pressure increases, the mean quenching wall heat flux ${\Phi_{\mathrm{q,2D}}} / \Phi_{\mathrm{q,1D}}$ rises, while the relative quenching distance ${x_{\mathrm{q,2D}}} / {x_{\mathrm{q,1D}}}$ decreases.
    However, the relative increase in wall heat flux only reaches values of up to $1.4$.
    Also for this parametric variation, the trend is consistent with the increasing intensity of thermodiffusive instabilities with an increasing pressure.
\end{itemize}

The analysis highlights the extent to which instabilities influence the wall heat flux and thus also the heat load on the combustor walls. 
This underscores the finding from part I \cite{PartI} that the analysis of HOQ in one-dimensional configurations is inadequate for operating conditions that are prone to thermodiffusive instabilities.
Furthermore, in cases where the flame front is only characterized by hydrodynamic instabilities and thermodiffusive instabilities are not significant, the mean value of the relative quenching wall heat flux is approximately $1$, with a standard deviation of approximately $0$, consistent with the corresponding one-dimensional case.
This leads to the conclusion that hydrodynamic instabilities have a negligible effect on the quenching process.

Since the analysis revealed that the trends of the increasing and decreasing relative quenching wall heat flux $\Phi_{\mathrm{q,2D}}/\Phi_{\mathrm{q,1D}}$ and quenching distance $x_{\mathrm{q,2D}}/x_{\mathrm{q,1D}}$ align with the increasing or decreasing intensities of thermodiffusive instabilities for all parametric variations, this relationship is further explored and quantified in the following. 

In Part I of this work \cite{PartI}, it was shown through varying the numerical domain sizes for a specific operating condition — and thereby altering the cell size distribution of the instabilities — that the mean relative quenching wall heat flux (two-dimensional unstable HOQ compared to one-dimensional stable HOQ) is linked to the intensity of thermodiffusive instabilities, quantified by the reactivity factor $I_0$.
Here, it is investigated whether this relationship is also applicable to the varying intensities of thermodiffusive instabilities due to different operating conditions.
\begin{figure}
    \centering
    \includegraphics{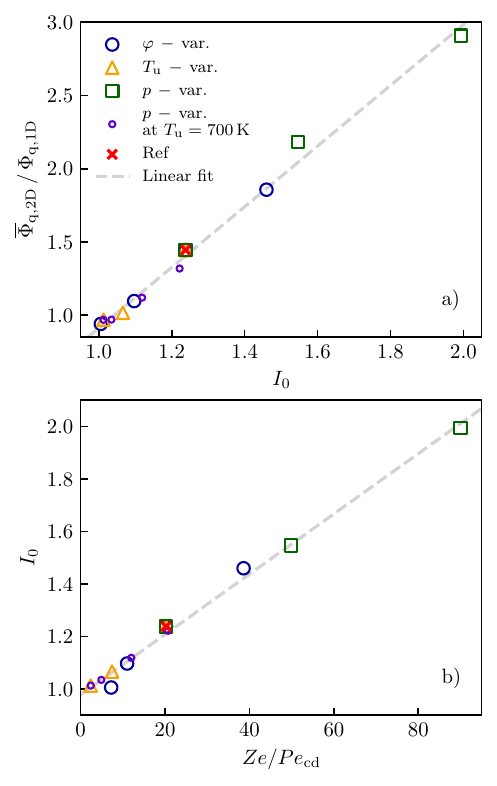}
    \caption{ \footnotesize
        a) Relative increase in the mean quenching wall heat flux $\overline{\Phi}_{\mathrm{q,2D}}  / \Phi_{\mathrm{q,1D}}$ (2D compared to 1D) over the reactivity factor $I_0$ for all parametric variations.
        b) Reactivity factor $I_0$ over the Zeldovich number divided by the convection-diffusion Péclet number $Ze/Pe_{\mathrm{cd}}$ of a one-dimensional freely-propagating flame at the identical operating condition for all parametric variations.
        The dashed grey line denotes a linear fit for the data points across all parametric variations.
        The red dot highlights the reference case `Ref`.
    }
    \label{fig:whf_over_I0}
\end{figure}
Figure \ref{fig:whf_over_I0} a) shows the mean relative quenching wall heat flux $\overline{\Phi}_{\mathrm{q,2D}} / \Phi_{\mathrm{q,1D}}$ over the reactivity factor $I_0$ of the respective two-dimensional freely-propagating flame for the whole parametric space.
Also in this case of varying operating conditions, an increase in reactivity (indicated by $I_0$) associated with thermodiffusive instabilities (see also \cite{Berger2023}) results in a higher mean quenching wall heat flux compared to the respective one-dimensional HOQ, where instabilities are inherently suppressed.
Furthermore, the data points for each parametric variation follow a linear relationship.
Notably, a linear relationship also holds across all operating conditions, as indicated by the linear fit over the entire dataset.

This figure extends the insights from part I across varying operating conditions, revealing that the intensified wall heat flux linked to instabilities is driven by the enhanced local reactivity due to local effects of thermodiffusive instabilities on the flame structure, while kinematic effects play a subordinate role in the quenching process.
This is further supported by the observation that purely hydrodynamic instabilities have no significant impact on the quenching process (see, for instance, the cases `phi1.0` and `Tu700` in Fig. \ref{fig:HOQ_over_domain}), even though their flame surface area $A/L_y$ and therefore the global consumption speed $s_{\mathrm{c}}/s_{\mathrm{l}}$ is enlarged, while the reactivity factor remains $I_0 = 1$.

The discussion of Fig. \ref{fig:HOQ_over_domain} and Fig. \ref{fig:whf_and_quenching_distance_2D_to_1D} further suggests that global quantities of the FWI can be estimated using the reactivity factor $I_0$.
To facilitate such an approximation for a specific mixture at a given operating condition, without the need for an extensive computation of a two-dimensional unstable freely-propagating flame, the following components are necessary:
\begin{itemize}[noitemsep]
    \item an approximation of the mean relative quenching wall heat flux $\overline{\Phi}_{\mathrm{q,2D}}/\Phi_{\mathrm{q,1D}}$ as a function of the reactivity factor $I_0$,
    \item and an approximation of the reactivity factor $I_0$ itself, based on key quantities of a one-dimensional flame for the respective operating condition.
\end{itemize}

The linear fit in Fig. \ref{fig:whf_over_I0} a) between the relative increase in the mean quenching wall heat flux $\overline{\Phi}_{\mathrm{q,2D}}/\Phi_{\mathrm{q,1D}}$ and the reactivity factor $I_0$ is given as:
\begin{equation}
    \overline{\Phi}_{\mathrm{q,2D}}/\Phi_{\mathrm{q,1D}} = a + b \cdot I_0 \, , 
\end{equation}
with the offset $a=-1.16$ and the slope $b = 2.07$.

To estimate the relative increase in the mean quenching wall heat flux from key quantities of a one-dimensional freely-propagating flame, a model for the reactivity factor $I_0$ is suggested.
Following the approach of Rieth et al.~\cite{Rieth2023}, who proposed a linear model for the reactivity factor $I_0$ as a function of the ratio between the Zeldovich number and the convection-diffusion Péclet number ratio $Ze/Pe_{\mathrm{cd}}$ for laminar flames, the reactivity factor $I_0$ is plotted against the ratio $Ze/Pe_{\mathrm{cd}}$ for all operating conditions in Fig. \ref{fig:whf_over_I0} b).
An increase in the Zeldovich number $Ze$ and a decrease in the convection-diffusion Péclet number $Pe_{\mathrm{cd}}$ result in a higher reactivity factor $I_0$ as a result of the enhanced intensity of thermodiffusive instabilities.

Consequently, the following linear model fit is employed:
\begin{equation}
    I_0 = c + d \cdot Ze / Pe_{\mathrm{cd}} \, , 
    \label{eq:lin_fit_Ze_Pe}
\end{equation}
with the offset $c=0.978$ and the slope $d=0.0115$. 
The Zeldovich number $Ze$ primarily characterizes the flame's reactivity, while the convection-diffusion Péclet number $Pe_{\mathrm{cd}}$ represents its tendency to become thermodiffusively unstable \cite{Rieth2023}.
It should be noted that alternative parameters to the convection-diffusion Péclet number $Pe_{\mathrm{cd}}$, which provide different measures of the relative importance of molecular diffusion in promoting thermodiffusive instabilities, such as the effective Lewis number $Le_{\mathrm{eff}}$, can also be employed for scaling, as shown in the fit of the normalized global consumption speed $s_{\mathrm{c}}/s_{\mathrm{l}}$ \cite{Berger2022, Berger2022a}.
Unlike in \cite{Berger2022, Berger2022a}, however, the expansion ratio $\sigma = \rho_{\mathrm{u}} / \rho_{\mathrm{b}}$ plays a secondary role here, as it primarily captures the additional effects of hydrodynamic instabilities and therefore contributes to the surface area increase $A/L_y$, but not to the increased reactivity factor $I_0$ \cite{Berger2023}.

\begin{figure}
    \centering
    \includegraphics{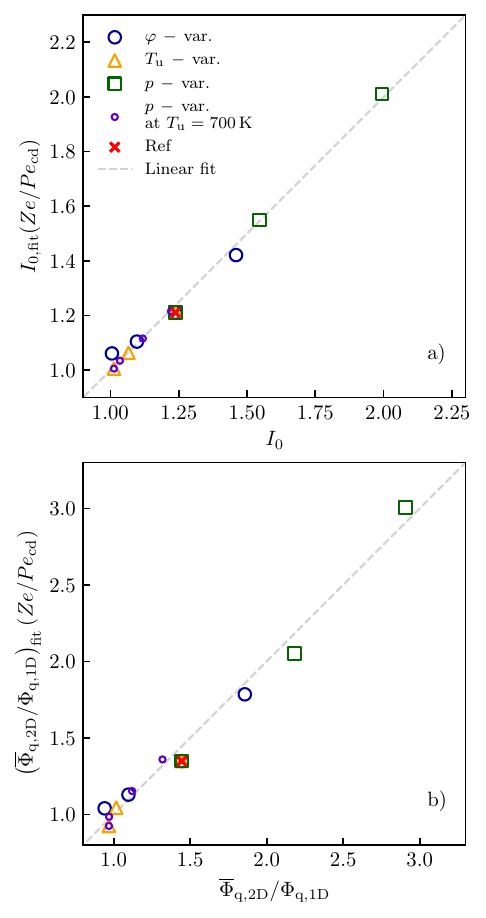}
    \caption{\footnotesize
        a): Linear model fit based on the ratio of the Zeldovich number and the convection-diffusion Péclet number $Ze/Pe_{\mathrm{cd}}$ (Eq. \ref{eq:lin_fit_Ze_Pe}): Comparison of the numerically measured reactivity increase $I_0$ versus the predicted values by Eq. \ref{eq:lin_fit_Ze_Pe}
        b): Combined model approximation for the relative increase in the mean quenching wall heat flux $\overline{\Phi}_{\mathrm{q,2D}}/\Phi_{\mathrm{q,1D}}$ as a function of the ratio of the Zeldovich number and the convection-diffusion Péclet number $Ze/Pe_{\mathrm{cd}}$ (Eq. \ref{eq:combined_fit_whf}): Comparison of the numerically measured values versus the predicted values by Eq. \ref{eq:combined_fit_whf}.
    }
    \label{fig:I0_model}
\end{figure}

Figure \ref{fig:I0_model} a) shows a comparison of the numerically measured values of the reactivity factor $I_0$ and the values from the model (Eq. \ref{eq:lin_fit_Ze_Pe}).
A strong agreement between the numerically measured and predicted values from the model equation is observed.

Building on these two model approximations, the following combined model approximation is proposed for the relative increase in the mean quenching wall heat flux $\overline{\Phi}_{\mathrm{q,2D}}/\Phi_{\mathrm{q,1D}}$ as a function ratio between the Zeldovich number and the convection-diffusion Péclet number ($Ze/Pe_{\mathrm{cd}}$):
\begin{equation}
    \overline{\Phi}_{\mathrm{q,2D}}/\Phi_{\mathrm{q,1D}} = a + b \cdot \left(c + d \cdot Ze/Pe_{\mathrm{cd}} \right) = e + f \cdot Ze/Pe_{\mathrm{cd}} \, ,
    \label{eq:combined_fit_whf}
\end{equation}
with the offset $e=0.864$ and the slope $f=0.0238$. 
Figure \ref{fig:I0_model} b) shows a comparison of the numerically measured values of the relative increase in the mean wall heat flux $\overline{\Phi}_{\mathrm{q,2D}}/\Phi_{\mathrm{q,1D}}$ and the values from the joint model approximation (Eq. \ref{eq:combined_fit_whf}).
For this joint model as well, a strong agreement is observed between the numerically measured values and those predicted by the model equation.
Thus, the absolute value of the mean quenching wall heat flux $\overline{\Phi}_{\mathrm{q,2D}}$ of an intrinsically unstable flame can be estimated based on a one-dimensional freely-propagating flame and a one-dimensional head-on quenching simulation.

\section{Conclusions} \addvspace{10pt}
\label{sec:Conclusions}

The present study investigates the interaction of intrinsically unstable hydrogen/air flames with combustor walls in a two-dimensional head-on quenching configuration.
Direct numerical simulations over a wide range of equivalence ratios $\varphi = 0.4 - 1$, unburnt gas temperatures $T_{\mathrm{u}} = \SI{298}{\kelvin} - \SI{700}{\kelvin}$, and pressures $p = \SI{1.01325}{\bar}- \SI{20}{\bar}$ are performed.

In the first step, one-dimensional freely-propagating and head-on quenching flames -- where instabilities are inherently suppressed -- are analyzed.
These flames characterize the effect of the changing operation conditions across the parametric space and provide reference values for the subsequent analysis of the unstable, two-dimensional quenching processes.

Subsequently, two-dimensional unstable hydrogen/air flames are assessed.
The investigated parameter space includes flames with varying intensities of hydrodynamic and thermodiffusive instabilities, allowing for an assessment of the sensitivity of quenching characteristics on the intensity of these instabilities.
The parametric variations reveal a negligible influence of hydrodynamic instabilities on key quantities of the quenching process. 
In contrast, the intensity of thermodiffusive instabilities is directly linked to an increase in both the average and local wall heat flux compared to the one-dimensional flames under identical operating conditions.
Given the investigated parameter space, this implies that: 
\begin{itemize} [noitemsep] 
    \item \textbf{Variation of $\varphi$ and $T_{\mathrm{u}}$:} Lower values of the equivalence ratio $\varphi$ or the unburnt gas temperature $T_{\mathrm{u}}$ increase the intensity of thermodiffusive instabilities, resulting in higher wall heat fluxes. 
    \item \textbf{Variation of $p$:} Higher values of the pressure $p$ increase the intensity of thermodiffusive instabilities, leading to an increased wall heat flux compared to the head-on quenching of a one-dimensional stable flame under identical operating conditions.
\end{itemize}

Across the entire parameter space, the intensity of thermodiffusive instabilities can be captured by the flame reactivity, quantified by the reactivity factor $I_0$. 
Combined with an additional model fit for the reactivity factor $I_0$ -- based on key parameters of a one-dimensional freely-propagating flame under the respective operating conditions -- a joint model is proposed for estimating the thermal load on combustor walls using only key quantities of the one-dimensional freely-propagating flame and head-on quenching.

\newpage

\section*{Disclosure statement}
The authors declare that they have no known competing financial interests or personal relationships that could have appeared to influence the work reported in this paper.

\section*{Acknowledgements}
This work has been funded by the Deutsche Forschungsgemeinschaft (DFG, German Research Foundation) -- Project Number 523792378 -- SPP 2419.
The simulations were performed on the Lichtenberg high-performance computer at TU Darmstadt.

\newpage

\appendix

\section{Global quenching properties for one-dimensional and two-dimensional HOQ}
\label{sec:app_add_reps}

\begin{figure}[H]
    \centering
    \includegraphics[scale=0.8]{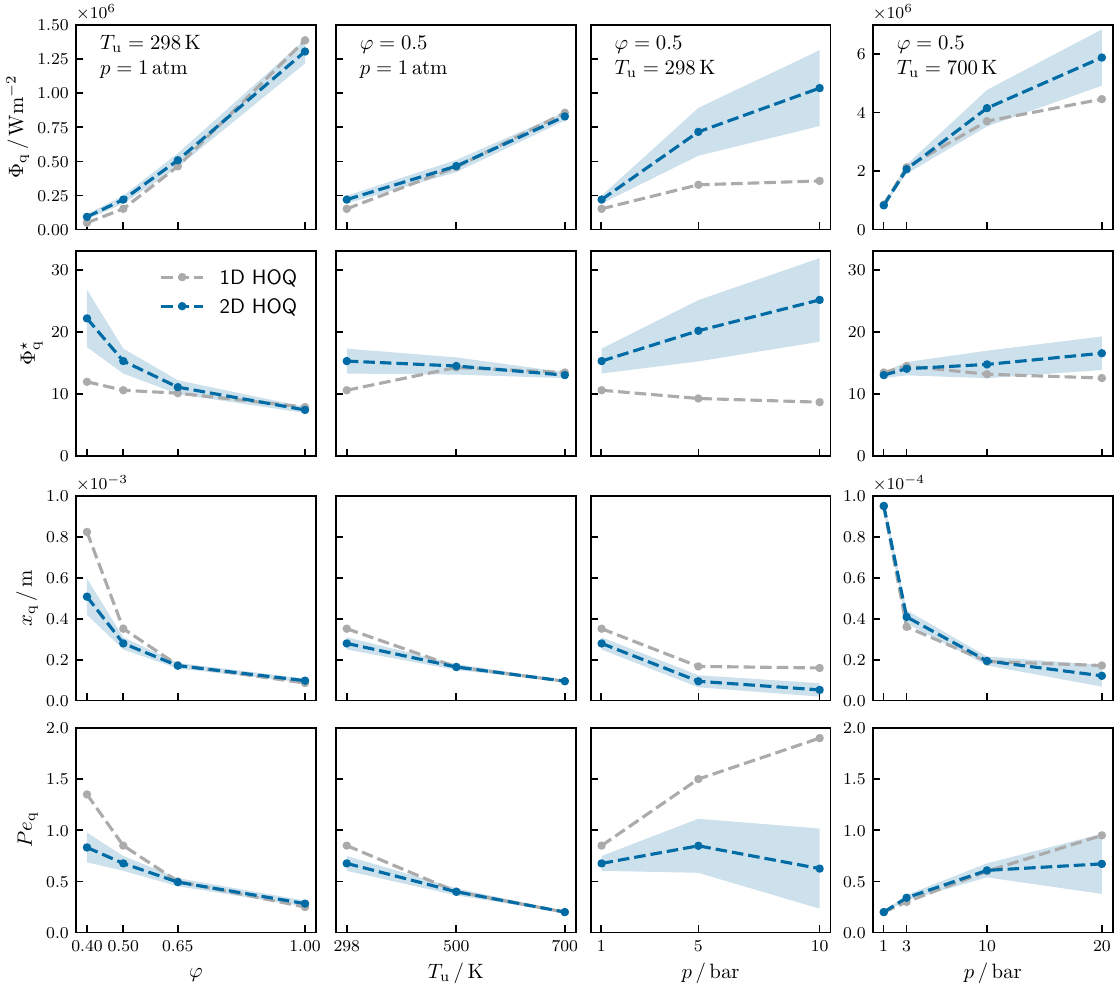}
    \caption{ \footnotesize
        Quenching wall heat flux $\Phi_{\mathrm{q}}$ (first row), normalized quenching wall heat flux $\Phi^{\star}_{\mathrm{q}}$ (second row), quenching distance $x_{\mathrm{q}}$ (third row) and quenching Péclet number $Pe_{\mathrm{q}}$ (fourth row) for the one-dimensional and the two-dimensional HOQs (mean value $\pm 1$ standard deviation $\sigma$) for the parametric variations of the equivalence ratio $\varphi$, unburnt temperature $T_{\mathrm{u}}$ and the pressure $p$.      
    }
    \label{fig:overview_whf_qd_norm_and_nonnorm}
\end{figure}

\newpage

\bibliographystyle{unsrtnat_mod}
\bibliography{publication.bib}

\end{document}